\newif\ifnotblinded
\notblindedtrue  

\documentclass[a4paper,12pt]{article}

\usepackage[natbibapa]{apacite}
\usepackage{amsmath}
\usepackage{mathtools}
\usepackage{setspace}
\usepackage{placeins}
\usepackage{appendix}
\usepackage{minibox}
\usepackage{graphicx}
\usepackage{tikz}
\usepackage{xcolor}
\usepackage{natbib}
\usepackage{caption}
\usepackage{booktabs}
\usepackage{multirow}
\usepackage{authblk}
\usepackage{epstopdf}
\usepackage{algpseudocode}
\usepackage{float}
\usepackage[utf8x]{inputenc}
\usepackage{amsfonts}
\usepackage{latexsym}
\usepackage{adjustbox}
\usepackage{array}
\usepackage{lscape}
\usepackage{amssymb}
\usepackage[hmargin = 3cm, vmargin = 2.5cm]{geometry}

\usepackage{xcolor,colortbl}

\usepackage[column=0]{cellspace}
\newcommand{\cred}{\textcolor{red}}
\newcommand{\cg}{\cellcolor[HTML]{C0C0C0}}
\newcommand{\cb}{\color[HTML]{3166FF}}
\newcommand{\fb}{\framebox[0.08\columnwidth]}
\newcommand{\E}{\mathop{\mathbb E}}

\newcommand{\utwi}[1]{\mbox{\boldmath $ #1$}}

\begin{document}

\title{
\bf Semi-parametric financial risk forecasting incorporating multiple realized measures}

\ifnotblinded
\author{Rangika Peiris}
\author{Chao Wang}
\author{Richard Gerlach}
\author{Minh-Ngoc Tran\thanks{Email for all authors: \{rangika.peiris, chao.wang, richard.gerlach, minh-ngoc.tran\}@sydney.edu.au}}
\affil{Discipline of Business Analytics, The University of Sydney}
\date{} \maketitle
\begin{abstract}
A semi-parametric joint Value-at-Risk (VaR) and Expected Shortfall (ES) forecasting framework employing multiple realized measures is developed. The proposed framework extends the realized exponential GARCH model to be semi-parametrically estimated, via a joint loss function, whilst extending existing quantile time series models to incorporate multiple realized measures. A quasi-likelihood is built, employing the asymmetric Laplace distribution that is directly linked to a joint loss function, which enables Bayesian inference for the proposed model. An adaptive Markov Chain Monte Carlo method is used for the model estimation. The empirical section evaluates the performance of the proposed framework with six stock markets from January 2000 to June 2022, covering the period of COVID-19. Three realized measures, including 5-minute realized variance, bi-power variation, and realized kernel, are incorporated and evaluated in the proposed framework. One-step-ahead 1\% and 2.5\% VaR and ES forecasting results of the proposed model are compared to a range of parametric and semi-parametric models, lending support to the effectiveness of the proposed framework.

\end{abstract}

\section{Introduction}

Financial risk management is an integral task for financial institutions. Value-at-Risk (VaR) is a standard tool for measuring and controlling financial market risks. Let $\mathcal{L}_{t}$ denote the information available at time $t$ and 
\begin{center}
  $F_{t}(r)=Pr(r_t \leq r \mid \mathcal{L}_{t-1})$  
\end{center}
be the Cumulative Distribution Function (CDF) of the return $r_{t}$ conditional on $\mathcal{L}_{t-1}$. Assuming that $F_{t}(\cdot)$ is strictly increasing and continuous on the real line $\mathbb{R}$, the one-step-ahead $\alpha$-level Value-at-Risk (VaR) at time $t$ can be defined as:
\begin{center}
  $  Q_{t} = F_{t}^{-1}(\alpha),   \quad 0 < \alpha < 1. $
\end{center}

However, VaR cannot measure the magnitude of the loss for violations and is not mathematically coherent, meaning that it is not a sub-additive measure and can favour non-diversification. \cite{artzner1999coherent}  propose an alternative called Expected Shortfall (ES), also called conditional VaR or tail VaR. ES calculates the expected loss conditional on exceeding a VaR threshold and is a coherent risk measure. The one-step-ahead $\alpha$-level ES is the tail conditional expectation of $r_{t}$, i.e.: 
 \begin{center}
     $\textup{ES}_{t}= E(r_{t}\mid r_{t}\leq Q_{t,}\mathcal{L} _{t-1}).$
 \end{center}
 
The recent Basel III Accord \citep{BIS2023} places new emphasis on ES. Our paper focuses on the daily forecasting of VaR and ES on the lower/left tail. Following the Basel III Accord, the common $\alpha= 2.5\%$ probability level is studied in the paper. The more extreme 1\% probability level is also investigated.

Volatility plays a crucial role in parametric tail risk forecasting. The GARCH model of \cite{engle1982autoregressive} and \cite{bollerslev1986generalized} is widely used for modelling and forecasting volatility in the finance industry. Numerous extensions, such as EGARCH by \citet{nelson1991conditional} and GJR-GARCH by \citet{glosten1993relation}, have been introduced to capture the well-known leverage effect. However, the volatility dynamics in these conventional GARCH models is driven by (daily) returns, which are considered potentially noisy signals for the volatility series.

The availability of high-frequency intra-day data has allowed the construction of many informative, efficient realized measures (RMs) of volatility. The most commonly used RMs include Realized variance (RV) (\cite{andersen1998answering}, \cite{andersen2003modeling}), Realized Range (RR) (\cite{christensen2007realized}, \cite{martens2007measuring}), Realized Kernel (RK) (\cite{barndorff2009realized}, and Bi-power variation (BV) (\cite{barndorff2004power}), etc.

\cite{hansen2012realized} include a RM in their volatility equation via their realized GARCH framework, enabling joint modelling of returns and RMs using a measurement equation. \cite{hansen2016exponential} extend the realized GARCH framework to include multiple RMs via the Realized Exponential GARCH (REGARCH) model. The REGARCH shows improved volatility forecasting performance compared to realized GARCH, and GARCH, demonstrating the usefulness of incorporating multiple realized measures in volatility modelling.

The tail risk forecasting accuracy of parametric models depends heavily on the choice of the distribution of the returns. Semi-parametric models, which do not rely on a specific return distribution, are also developed in the literature. \cite{engle2004caviar} introduce the conditional auto-regressive VaR (CAViaR) model, which directly estimates VaR as the quantile of the conditional return distribution via a quantile regression framework. The model is optimized by minimising the quantile loss function. However, CAViaR models do not estimate ES. 

\citet{fissler2016higher} show that VaR and ES are jointly elicitable for a class of joint loss functions, although ES is not elicitable by itself. This finding carries significant implications for the risk forecasting literature, particularly for researchers in the field of semi-parametric risk forecasting, who have new avenues to explore in joint VaR and ES modelling. \citet{taylor2019forecasting} proposes an ES-CAViaR framework to jointly and semi-parametrically estimate VaR and ES. A quasi-likelihood, built on the asymmetric Laplace (AL) distribution, allows the joint estimation of conditional VaR and ES. \citet{taylor2019forecasting} shows that the AL quasi-likelihood function falls into the class of strictly consistent loss functions developed by \citet{fissler2016higher}. In the ES-CAViaR model, a CaViaR-type quantile equation models the VaR component. Then, ES is modelled via two proposed versions of a VaR to ES relationship: additive and multiplicative. 

\citet{gerlach2020semi} incorporate a realized measure as an exogenous variable, extending ES-CAViaR models to the semi-parametric ES-X-CAViaR-X model class, finding improved VaR and ES forecast performance. \cite{wang2023semi} further extend the work of \citet{gerlach2020semi} by introducing the semi-parametric Realized-ES-CAViaR framework, including a measurement equation to model the relationship between the tail risk measure and a RM; a leverage effect is also considered in the framework. 

Three key facts motivate the development of our proposed framework. First, the REGARCH, using multiple RMs to model volatility, demonstrates improved performance compared to the realized GARCH, using only one RM. Second, the REGARCH is a parametric model requiring the specification of a return error distribution, while semi-parametric models, such as ES-CAViaR, do not, which is advantageous in many real return series. Third, incorporating a single RM into the semi-parametric modelling process, such as the ES-X-CAViaR-X or Realized-ES-CAViaR, can improve risk forecasting accuracy. Therefore, there is a gap in the literature regarding semi-parametric joint VaR and ES forecasting models with multiple realized measures; filling that gap is the primary aim of this paper.

The main contributions of this paper are as follows. First, a new semi-parametric joint VaR and ES forecasting framework incorporating multiple RMs is proposed. This extends the quantile regression framework using multiple RMs as exogenous variables. The relationship between VaR and ES is modelled as time-varying and driven by the information from RMs. Further, a measurement equation is included in the framework to model the joint contemporaneous dependencies between the quantile series and multiple RMs. Second, an adaptive Bayesian MCMC algorithm is used to estimate the proposed model, including the parameters in the measurement equation variance-covariance matrix. Lastly, the effectiveness of the proposed framework is evaluated via a comprehensive empirical study, including 33 competing models and covering the period from January 2000 to June 2022. The code of implementing the proposed model publicly available at: 
\\
\url{https://github.com/chaowang-usyd/Realized-ES-CAViaR-M}.

This paper is organized as follows. Section 2 reviews the relevant existing literature on tail risk forecasting models. Section 3 presents the proposed framework. The likelihood function and the adaptive Bayesian MCMC algorithm are presented in Section 4. Section 5 presents the empirical results. Section 6 concludes the paper. 

\section{Background models}\label{background_models_section}

This section describes the relevant models used to forecast VaR and ES in the literature, while the properties of each model are described in the context of motivating the proposed framework. Fundamental concepts used in the model development process are also discussed.

\subsection{Parametric GARCH-type models}\label{garch_section}

Let $\mathbf{r}=\left\{r_t, t=1, \ldots, T\right\}$ be a time series of daily returns. The key interest in parametric volatility modelling is the conditional variance, $\sigma_t^2=\operatorname{var}\left(r_t \mid \mathcal{L}_{t-1}\right)$. 
$\sigma_{t}$ is called the volatility. Here, $\E(r_t|\mathcal{L}_{t-1})=0$ is assumed, equivalent to working with demeaned returns in practice. The GARCH(1,1) model is:
\begin{center}
    $r_{t}=\sigma_{t}z_{t},$ \\
    $\sigma^2_{t} = \omega_{0} + \alpha_{1}r^2_{t-1} + \beta_{1}\sigma^2_{t-1}, $
\end{center}
where $z_t$ is i.i.d. with zero mean and unit variance. 
Parametric approaches require a parametric distribution for $z_t$ to be chosen, e.g., Gaussian or Student's $t$, to compute VaR and ES forecasts.

\citet{francq2015risk} and \citet{gao2008estimation} consider a semi-parametric approach using historical simulation (HS), by modelling VaR and ES as constant multiples of the latent volatility $\sigma_{t}$ which is assumed to follow a GARCH-type volatility model. Assuming a constant conditional return distribution with zero mean, the VaR and ES are modelled as:
\begin{align} 
  Q_{t} &=a_{\alpha}\sigma_{t}; \: \: \text{ES}_{t}=b_{\alpha}\sigma_{t}; \: \: \frac{\text{ES}_{t}}{Q_{t}}=\frac{b_{\alpha}}{a_{\alpha}}>1,   \label{eq: 1}  
\end{align}
where $a_{\alpha}$ and $ b_{\alpha}$ are constant depending on the return distribution and can be estimated via HS on the standardized residuals $\frac{r_t} {\hat{\sigma}_{t}}$. The series $\hat{\sigma}_{t}$ is estimated first using quasi-maximum likelihood (QML) \citep{gao2008estimation}. 

\subsection{Realized (E)GARCH model}

A parametric realized GARCH framework, incorporating a RM into the volatility modelling process via a measurement equation, is developed in \citet{hansen2012realized}. \citet{hansen2016exponential} further extend the realized GARCH by incorporating multiple RMs and propose the parametric realized EGARCH (REGARCH) model. A log-REGARCH specification is defined as:
\begin{align}  \label{Realized_EGARCH_model}  
r_{t}&=\sigma_{t}z_{t}, \\  
\mathrm{log}(\sigma_{t})&= \omega+\beta \mathrm{log}(\sigma_{t-1})+\tau_{1}z_{t-1}+ \tau_{2}\left(z_{t-1}^{2}-1\right)+ \boldsymbol{\gamma}^{T}\mathbf{u}_{t-1}, \notag \\
\mathrm{log}(x_{j,t})&=\xi_{j}+\varphi_{j}\mathrm{log}(\sigma_{t})+\delta_{j,1}z_{t} +\delta_{j,2}(z_{t}^{2}-1) +{u}_{j,t}, \,\,  j = 1, 2, ..., K. \notag    
\end{align}
The three log-REGARCH equations, in order, are the return equation, the GARCH or volatility equation, and the measurement equation, respectively. The measurement equation defines the contemporaneous relationship between the (ex-post) RMs of volatility and the (ex-ante) volatility. Here, $K$ denotes the number of RMs, and $K=1$ defines the original realized GARCH model. $\mathbf{x}_{t} = (x_{1,t},...,x_{K,t})^T$ is the vector of RMs, at time $t$, on the same scale as $\sigma_{t}$, e.g., $\sqrt{RV}$. $z_t$ are i.i.d. with zero mean and unit variance and $\mathbf{u}_t = (u_{t,1},....)^T\sim N(0,\Sigma)$. 
$z_{t}$ and $\mathbf{u}_{t}$ are mutually and serially independent. Further, the coefficient $\boldsymbol{\gamma} = (\gamma_1, ..., \gamma_K)^T$ of $\mathbf{u}_{t-1}$ represents how
informative the RMs of day $t-1$ are about volatility on day $t$. 
The model uses two sets of leverage functions, both following the usual quadratic form, to model the leverage effect. 

\subsection{Semi-parametric ES-X-CAViaR-X model}\label{ES-X-CAViaR-X_Section}

\citet{taylor2019forecasting} proposes a semi-parametric class of models (called ES-CAViaR) to model the dynamics of VaR and ES jointly. \citet{gerlach2020semi} extend this model by adding various different ES to VaR relationships, allowing a single RM to influence both VaR and ES separately. One of their proposed semi-parametric ES-X-CAViaR-X models is defined as follows:
\begin{align}   \label{ES_X_CAViaR_X_model}
Q_{t}&=\beta _{0}+\beta _{1} x_{t-1}+\beta _{2}Q_{t-1}, \\
\omega_{t}&=\gamma_{0} + \gamma_{1}x_{t-1}+ \gamma_{2}\omega_{t-1}, \notag \\  
\text{ES}_{t}&=Q_{t}-\omega_{t}.  \notag
\end{align}
Here $x_{t}$ is the RM. The dynamics of $\text{ES}_{t}$ and $Q_{t}$ have an additive, time-varying relationship, defined by $\omega_{t}$, which is driven separately to $Q_t$, by the RM. The $\omega_{t}$ specification is directly generalized from a GARCH-type model. This specification allows the unknown conditional return distribution to change over time.
The restriction $\gamma_{0} \geq 0, \gamma_{1} \geq 0, \gamma_{2} \geq 0$ is employed to ensure that the VaR and ES series do not cross.

\subsection{Semi-parametric Realized-ES-CAViaR models}\label{Re-ES-CAViaR_Section}

The semi-parametric Realized-ES-CAViaR models \citep{wang2023semi} extend the ES-X-CAViaR-X model by incorporating a measurement equation:

Non-crossing of VaR and ES is enforced via the condition $\gamma_{0} \geq 0, \gamma_{1} \geq 0, \gamma_{2} \geq 0$. The multiplicative error term $\epsilon_t= \frac{r_t} {Q_t}$ in the measurement equation facilitates the incorporation of the leverage effect into the model. Furthermore, the influence of the RM on VaR and ES is individually captured through the difference $\omega_{t}$. As the relationship between VaR and ES varies over time, driven by the RM, the conditional return distribution also evolves. Compared to the ES-X-CAViaR-X model \citep{gerlach2020semi}, the added measurement equation ``completes'' the model by regressing the RM on the quantile (can also be replaced with ES).

\section{Proposed Model}\label{proposed_model_section}
This paper proposes a new Realized-ES-CAViaR-M model, employing multiple RMs to jointly and semi-parametrically model VaR and ES. The model extends the Realized-ES-CAViaR, via incorporating multiple RMs and adding a log specification, as well as the REGARCH, by virtue of being semi-parametric, i.e., the return distribution assumption is not required and the relationship between VaR and ES varies over time.

To motivate these proposals, now we present the process on developing the Realized-ES-CAViaR-M model. Given a REGARCH (model (\ref{Realized_EGARCH_model})) with a parametric return distribution $z_t$ in the return equation $r_{t}=\sigma_{t} z_{t}$, we have $\sigma_t= \frac{Q_t} {a_{\alpha}}$, meaning the quantile $Q_t$ is proportional to the volatility $\sigma_t$. The constant $a_{\alpha}$ is equal to the CDF inverse of the selected parametric return distribution. 

Since in our proposed model we do not assume the value of $a_{\alpha}$, no return distribution is assumed and we have a semi-parametric approach, as defined in equation (\ref{eq: 1}). Further, we define the multiplicative error $ \epsilon_{t} = \frac{r_t} {Q_t} $ as that in the Realized-ES-CAViaR, then we have $\epsilon_{t} = \frac{\sigma_{t} z_{t}} {Q_t} = \frac{z_{t}}{a_{\alpha}} $ . Substituting $\sigma_t= \frac{Q_t} {a_{\alpha}}$ and $z_t= \epsilon_{t} a_{\alpha}$ into the REGARCH framework and removing the return equation produce:
\begin{align}   
\mathrm{log}(\frac{Q_t} {a_{\alpha}} )&= \omega+\beta \mathrm{log}( \frac{Q_{t-1}} {a_{\alpha}})+\tau_{1} \epsilon_{t-1} a_{\alpha} + \tau_{2}\left(\epsilon_{t-1}^2 a_{\alpha} ^{2}-1\right)+ \boldsymbol{\gamma}^{T}\mathbf{u}_{t-1}, \label{derive_1_1} \\
\mathrm{log}(x_{j,t})&=\xi_{j}+\varphi_{j}\mathrm{log}(\frac{Q_t} {a_{\alpha}})+\delta_{j,1} \epsilon_{t} a_{\alpha} +\delta_{j,2}(\epsilon_{t}^2 a_{\alpha}^{2}-1) +{u}_{j,t}, \,\,  j = 1, 2, ..., K.  \label{derive_1_2}   
\end{align}
Here, both $Q_t$ and $a_{\alpha}$ have negative values, as the left tail 1\% and 2.5\% probability levels are considered. Multiplying both side of (\ref{derive_1_1}) with a constant $c=\mathrm{log}(-a_{\alpha})$ and rearranging equation (\ref{derive_1_2}), we have:
\begin{align} \label{Realized_EGARCH_derive_2}  
\mathrm{log}(-Q_t) &= c \omega+\beta \mathrm{log} (-Q_{t-1})+c \tau_{1} a_{\alpha} \epsilon_{t-1}  + c\tau_{2} a_{\alpha} ^{2} \epsilon_{t-1}^2 - c \tau_{2} + c\boldsymbol{\gamma}^{T}\mathbf{u}_{t-1}, \notag \\
\mathrm{log}(x_{j,t})&=\xi_{j} - \varphi_{j} c \mathrm{log}(-Q_t)+\delta_{j,1} a_{\alpha} \epsilon_{t}  +\delta_{j,2} a_{\alpha}^{2} \epsilon_{t}^2 - \delta_{j,2} +{u}_{j,t}, \,\,  j = 1, 2, ..., K.   
\end{align}
Setting $w^*= c \omega - c \tau_{2}, \tau_1^*= c \tau_{1} a_{\alpha}, \tau_2^* = c \tau_{2} a_{\alpha} ^{2}, \boldsymbol{\gamma}^{*T}= c \boldsymbol{\gamma}^{T}, \xi_{j}^*= \xi_{j}- \delta_{j,2}, \varphi_{j}^*= -\varphi_{j}c, \delta_{j,1}^* = \delta_{j,1} a_{\alpha}, \delta_{j,2}^*= \delta_{j,2} a_{\alpha}^{2}$, the resulting model is:
\begin{align} \label{Realized_EGARCH_derive_3}  
\mathrm{log}(-Q_{t})&=\omega^*+\beta\mathrm{log}{(-Q_{t-1})}+\tau_{1}^*\epsilon_{t-1}+\tau_{2}^*\epsilon_{t-1}^2+\boldsymbol{\gamma}^{*T}\mathbf{u}_{t-1},  \notag \\ 
\mathrm{log}(x_{j,t})&=\xi_{j}^*+ \varphi_{j}^* \mathrm{log}(-Q_{t})+\delta_{j,1}^*\epsilon_{t}+\delta_{j,2}^*\epsilon_{t}^2+ u_{j,t}; \,\, j = 1, 2, ..., K.
\end{align}

The current framework does not contain the ES component. \cite{wang2023semi} develop an additive VaR to ES time varying $w_t$ component which is directly driven by the realized measure. We extend the approach by developing a $\omega_t$ component that is also separately driven by multiple realized measures. Therefore, the proposed model is specified as:

\textbf{Realized-ES-CAViaR-M}
\begin{align}
\mathrm{log}(-Q_{t})&= \omega+\beta\mathrm{log}(-Q_{t-1})+\tau_{1}\epsilon_{t-1}+\tau_{2}\epsilon_{t-1}^2+\boldsymbol{\gamma}^{T}\mathbf{u}_{t-1}, \label{eq: 5} \\ 
\omega_{t}&=\nu_{0}+\nu_{1}\omega_{t-1}+ \boldsymbol{\psi}^{T} \mathbf{|{u}|}_{t-1}, \label{eq: 6}  \\ 
\text{ES}_{t}&= Q_{t}-\omega_{t}, \label{eq: 7} \\
\mathrm{log}(x_{j,t})&=\xi_{j}+ \varphi_{j} \mathrm{log}(-Q_{t})+\delta_{j,1}\epsilon_{t}+\delta_{j,2}\epsilon_{t}^2+ u_{j,t}; \,\, j = 1, 2, ..., K.  \label{eq: 8}
\end{align}
We omit the * in the derived model (\ref{Realized_EGARCH_derive_3}) for cleanness of the presentation. The model contains four equations: the \textit{quantile} equation (\ref{eq: 5}), the VaR-ES difference \textit{$\omega_{t}$} equation (\ref{eq: 6}), the \textit{ES} equation (\ref{eq: 7}) and the \textit{measurement} equations (\ref{eq: 8}). As in REGARCH, $K$ is the number of realized measures and $K=1$ gives a log specification of the Realized-ES-CAViaR. Here $x_{j,t}$ is the square root of the RM, i.e., on the same scale as volatility. The measurement error vector $\mathbf{u}_{t}\stackrel{\text{i.i.d.}}{\sim} N\left(0,\Sigma\right)$, as standard. $\Sigma$ is the variance-covariance matrix of $\mathbf{u}_{t}$, with dimension $K \times K$. The key developments of the model are now discussed.

The quantile equation (\ref{eq: 5}) extends the existing quantile regression (CAViaR) by introducing a log specification, including the leverage effect term and incorporating the information from multiple RMs. This makes the model analogous to REGARCH. This paper studies the left tail quantile, for example, $\alpha=1\%; 2.5\% $, where each quantile $Q_{t}$ is less than 0, leading to the utilization of $(-Q_{t})$ in the log operator. Incorporating a log specification also guarantees that $(-Q_{t})$ is always positive, thus $Q_{t}$ is guaranteed to be negative for the studied left tail. The quadratic leverage effect specification as in REGARCH is followed. The regression coefficients $\boldsymbol{\gamma}$ capture how influential the $K$ lagged RMs are on next period (log-)quantile. 

The $\omega_t$ equation  (\ref{eq: 6}) and ES equation (\ref{eq: 7}) capture the time-varying and additive relationship between VaR and ES, all driven separately by the lagged RM vector, whose individual effects on the VaR to ES difference are given by $\boldsymbol{\psi}^{T}$, where $\boldsymbol{\psi}= (\psi_1, ..., \psi_K)^T$. Again, we constrain $\nu_{0} \geq 0, \nu_{1} \geq 0, \boldsymbol{\psi} \geq 0$ to ensure that the VaR and ES series do not cross. Since the $\text{ES}_{t}$ is modelled as $Q_{t}$ minus a non-negative time varying $\omega_{t}$ component, the leverage effect included in the quantile equation is implicitly applied to ES as well.  Other relationships between VaR and ES, such as the multiplicative one in \citet{gerlach2020semi}, could also be explored.

The measurement equations (\ref{eq: 8}) complete the model by providing a way to model the joint contemporaneous dependence between the risk level and multiple RMs $x_{j,t}; j = 1, 2, ..., K$. The leverage function follows the quadratic form as in REGARCH.

Comparing to the Realized-ES-CAViaR model, the Realized-ES-CAViaR-M extends it via incorporating the information of multiple realized measures during the quantile and ES forecasting process. Further, a log specification is used in Realized-ES-CAViaR-M. As below, we also introduce the Log-Realized-ES-CAViaR which aims to investigate the impact of adding the log specification into the quantile regression and measurement equations. The model will be compared with other models in the empirical study.

\textbf{Log-Realized-ES-CAViaR}
\begin{align} \label{Log_Realized_ES_CAViaR_model}
\mathrm{log}(-Q_{t})&=\beta _{0}+\beta _{1} \mathrm{log}(x_{t-1})+\beta _{2} \mathrm{log} (-Q_{t-1}), \\
\omega_{t}&= \gamma_{0}+\gamma_{1}x_{t-1} + \gamma_{2} \omega _{t-1}, \notag  \\
\text{ES}_{t}&=Q_{t}-\omega_{t}, \notag \\ 
\mathrm{log}(x_{t}) &= \xi+\phi \mathrm{log}(-Q_{t})+\tau_{1}\epsilon_{t}+ \tau_{2}\left ( \epsilon^2_{t}-E(\epsilon^2) \right ) +u_{t}. \notag 
\end{align}

\ifnotblinded
\section{Likelihood and model estimation}
CAViaR-type models are typically estimated via minimising the quantile loss function, for which the latent quantile series is strictly consistent. \cite{engle2004caviar} suggest that solutions to the optimization of the CAViaR type models can be heavily dependent on the chosen initial values. \cite{li2023bayesian} demonstrate that maximum likelihood based quantile regression models are sensitive to initial conditions and advocate for using Bayesian approach as a more reliable alternative. With an intensive simulation study, \cite{wang2023semi} have shown that Bayesian estimator is more accurate than the AL based QML estimator regarding parameter estimation and risk forecasting. Further, to conduct statistical inference Bayesian method can conveniently help us capture the parameter estimation uncertainty, by calculating the 95\% credible intervals using the MCMC chains. Therefore, this paper also employs the Bayesian methods.

\subsection{Likelihood function for the proposed model}\label{likelihood_function}

As discussed in Sections \ref{ES-X-CAViaR-X_Section} and \ref{Re-ES-CAViaR_Section}, the ES-CAViaR, ES-X-CAViaR-X and Realized-ES-CAViaR models are semi-parametric. However, Bayesian methods typically employ a parametric distributional assumption to form a likelihood.

\citet{koenker1999goodness} note that the conventional quantile regression estimator is equivalent to an MLE based on the AL density, with a mode at the quantile. Discovering a specific link between $\text{ES}_t$ and a dynamic $\sigma_t$, for the AL distribution, \citet{taylor2019forecasting} further extends this result to produce the conditional density function:
\begin{equation}\label{al_likelihood}
f_t(r) = \frac{(\alpha-1)}{\text{ES}_t} \exp \left( \frac{(r_t-Q_t)(\alpha - I(r_t \leq  Q_t))}{\alpha \text{ES}_t}  \right), 
\end{equation} 
As shown in \citet{taylor2019forecasting}, the negative logarithm of this AL-based density function is strictly consistent for $Q_{t}$ and $\text{ES}_{t}$ jointly, meaning that it fits into the class of strictly consistent loss functions developed by \cite{fissler2016higher}. 

This density then allows a quasi-likelihood function to be built, given models for $Q_t$ and $\text{ES}_t$, assuming a zero mean return, thus allowing Bayesian methods to be employed. Since $r_{t}$ cannot follow an AL distribution with a mode at $Q_t$, the AL-based likelihood built on equation (\ref{al_likelihood}) is a quasi-likelihood function, whose mode coincides with the minimum of the joint loss function. The quasi-log-likelihood is then:
\begin{equation}\label{al_log_likelihood}
\ell(\mathbf{r};\utwi{\theta})=\sum_{t=1}^{T} \left ( \log \frac{(\alpha -1)}{\text{ES}_{t}} + \frac{(r_{t}-Q_{t})(\alpha-I(r_{t}\leq Q_{t}))}{\alpha \text{ES}_{t}} \right ),
\end{equation} 
where $\mathbf{r} = \left \{ r_{1},r_{2},...r_{T} \right \} $ and the parameter vector is $\utwi{\theta}$.

The full model likelihood also includes parts from the measurement equations. The AL-based return quasi-log-likelihood (\ref{al_log_likelihood}) combines with the likelihood for the RMs to produce the full quasi-log-likelihood for the proposed Realized-ES-CAViaR-M model:
\begin{equation}  
\ell(\mathbf{r},\mathbf{X};\utwi{\theta},\Sigma)= \ell(\mathbf{r};\utwi{\theta})+ \ell(\mathbf{X}|\mathbf{r};\utwi{\theta},\Sigma), \notag
\end{equation}
where $\mathbf{X}$ is the set of multiple RMs: $\{x_{1,t},x_{2,t}, \ldots, x_{K,t} \}$ and $\Sigma$ is the covariance matrix of the measurement errors $\mathbf{u}_{t}$.

Thus, the full quasi-log-likelihood of the proposed model can be written as:
\begin{equation} 
\begin{split}
\ell(\mathbf{r},\mathbf{X};\utwi{\theta},\Sigma)=& \sum_{t=1}^{n} \left ( \log\frac{(\alpha -1)}{\text{ES}_{t}} + \frac{(r_{t}-Q_{t})(\alpha-I(r_{t}\leq Q_{t}))}{\alpha \text{ES}_{t}} \right ) \\
    & -\frac{1}{2} \sum_{t=1}^{n}  \left ( k\log(2\pi )+\log(\left | \Sigma  \right |)+\mathbf{u}_{t}^{'}(\utwi{\theta})\Sigma^{-1}\mathbf{u}_{t}(\utwi{\theta})\right), \label{eq: 11}
\end{split}
\end{equation}

For any given value of $\utwi{\theta}$, \citet{hansen2016exponential} show that the RM based Gaussian likelihood yields the partial maximization concerning $\Sigma$ as: 
 $$\hat{\Sigma}(\utwi{\theta})=\frac{1}{n}\sum_{t=1}^{n}\mathbf{u}_{t}(\utwi{\theta}){\mathbf{u}_{t}(\utwi{\theta})^{'}},$$
 where they point out that $\mathbf{u}_{t}$ in the above equation depends on $\theta$, but does not depend on the covariance matrix $\Sigma$. Therefore, the maximization problem is simplified to finding $\textup{argmax}_{\utwi{\theta}}\ell(\mathbf{r},\mathbf{X};\utwi{\theta},\hat{\Sigma}{(\utwi{\theta}}))$ since  $$\sum_{t-1}^{n}\mathbf{u}_{t}^{'}(\utwi{\theta})\Sigma (\theta)^{-1}\mathbf{u}_{t}(\theta)=tr\left \{ \sum_{t=1}^{n}\hat{\Sigma}(\utwi{\theta})^{-1}\mathbf{u}_{t}(\utwi{\utwi{\theta}})\mathbf{u}_{t}^{'}(\utwi{\utwi{\theta}}) \right \}=nK$$ which does not depend on $\utwi{\theta}$. 
 
 From a Bayesian standpoint, the likelihood is an inverse Wishart distribution in $\Sigma$, which can then be integrated out from the likelihood, as discussed in the next section. 

\subsection{Bayesian estimation}

The quasi-log-likelihood in (\ref{eq: 11}) includes the logarithm of the multivariate Gaussian density for each measurement error vector $\mathbf{u}_{t}$. Denoting $\utwi{\theta}_{-\Sigma}$ as the vector of all model parameters excluding $\Sigma$, the integrated measurement likelihood is:
\begin{align*}
    p(\mathbf{X} | \mathbf{r}, \utwi{\theta}_{-\Sigma}) =& \int p(\mathbf{X} | \mathbf{r}, \utwi{\theta} ) p(\Sigma) d \Sigma \\
    =& \int  (2\pi )^{-0.5 KT} \left | \Sigma  \right |^{-0.5T} \exp \left[-0.5 \sum_{t=1}^T \mathbf{u}_{t}^{'}\Sigma^{-1}\mathbf{u}_{t} \right]  p(\Sigma) d \Sigma  \,\, . 
\end{align*}
Under a standard Jeffreys prior $p(\Sigma) \propto |\Sigma|^{-0.5(K+1)}$, the integration is proportional to an inverse Wishart density function in $\Sigma$, so the integral can be shown to be:
\begin{align}\label{int_meas}
    p(\mathbf{X} | \mathbf{r}, \utwi{\theta}_{-\Sigma}) \propto & \, |\hat{\Sigma}|^{-0.5(T-K-1)}   \,\, ,
\end{align}
where the terms not in $\Sigma$ are ignored, and $\hat{\Sigma} = \frac{1}{T-K-1} \sum_{t=1}^T \mathbf{u}_{t}\mathbf{u}_{t}^{'}$ is the usual sample variance covariance estimator of $\Sigma$. Thus, the full integrated quasi-likelihood, replacing the log full measurement likelihood in (\ref{eq: 11}) with its integrated version in (\ref{int_meas}), logged, is:
\begin{equation} 
\begin{split}
\ell(\mathbf{r},\mathbf{X};\utwi{\theta}_{-\Sigma})=& \sum_{t=1}^{n} \left ( \log\frac{(\alpha -1)}{\text{ES}_{t}} + \frac{(r_{t}-Q_{t})(\alpha-I(r_{t}\leq Q_{t}))}{\alpha \text{ES}_{t}} \right ) \\
    & -\frac{1}{2} (T-K-1) \log(|\hat{\Sigma}|) \, . \label{eq: 13a}
\end{split}
\end{equation}

Priors are chosen to be flat over the regions sufficient for non-negativity of $\omega_{t}$ in equation (\ref{eq: 6}), combined with the others, quite liberal and wide, limits to ensure finite parameter ranges and a proper prior. Thus, we choose $\pi(\utwi{\theta}_{-\Sigma})\propto I(A)$, being a flat prior for $\utwi{\theta}_{-\Sigma}$ over the region $A$, and 0 elsewhere. To ensure finite parameter ranges, $A$ restricts each element of $\utwi{\theta}_{-\Sigma}$ to be inside $(-D_0,D_0)$. For example, stationarity requires $|\beta_1|<1$, i.e., $D_0 = 1$ for $\beta_1$. In the empirical study, we choose $ D_0 = 3$ for the other parameters, which is sufficiently large based on our analyses. To ensure non-negativity of $\omega_{t}$, region $A$ further restricts $\nu_{0} \geq 0, \nu_{1} \geq 0, \boldsymbol{\psi} \geq 0$. These priors take a log transformation and are combined with the integrated quasi-log-likelihood in equation (\ref{eq: 13a}) to construct the posterior distribution.

Following \citet{chen2022dynamic}, in the MCMC algorithm, to assist with the speed of mixing, the parameter vector is simulated in blocks, i.e., each block of parameters is simulated from its conditional posterior. Blocks are chosen so that parameters within each block tend to be more correlated in the posterior, whilst parameters not in the same block are less correlated; this aids in faster mixing and convergence. Table \ref{table: 1} details the blocking structure, based on the number of RMs in the model. The block-wise proposals are generated and accepted with the usual Metropolis algorithm, e.g., see \citet{chen2022dynamic}.

\begin{table}[H]
\centering
\renewcommand{\arraystretch}{0.5}
	\caption{Block structure of the employed MCMC. (Parameters in the variance-covariance matrix $\Sigma$ have been integrated out. $B_{i}$ represents the  $i^{th}$ parameter block)}
	\label{table: 1}
	\centering
	\begin{tabular}{ccccc}
		\hline
		 \textbf{Block number} & \textbf{k=1} (12 parameters) & \textbf{k=2}  (18 parameters) & \textbf{k=3}  (24 parameters)\\
		\hline
		$B_{1}$ &  $\left \{ \omega,\beta,\tau_{1}, \tau_{2}\right \}$   & $\left \{ \omega,\beta,\tau_{1}, \tau_{2}\right \}$  &   $\left \{ \omega,\beta,\tau_{1}, \tau_{2}\right \}$     \\
		
     	$B_{2}$ & $\left \{ \gamma, \delta_{11},\delta_{12} \right \}$	& $\left \{ \gamma_{1},\gamma_{2}, \xi_{1},\xi_{2}   \right \}$ &  $\left \{ \gamma_{1},\gamma_{2},\gamma_{3}  \right \}$    \\
		
		$B_{3}$ & $\left \{ \nu_{0},\nu_{1} \right \} $  & $\left \{ \varphi_{1} ,\varphi_{2}  \right \}$   &  $\left \{ \xi_{1},\xi_{2},\xi_{3}  \right \}$   \\
		
		$B_{4}$ & $\left \{ \xi,\varphi,\psi  \right \}$ &$\left \{ \delta_{11},\delta_{12},\delta_{21},\delta_{22}  \right \}$ &  $\left \{ \varphi_{1},\varphi_{2} ,\varphi_{3} \right \}$   \\
		
		$B_{5}$ & & $\left \{ \nu_{0},\nu_{1} \right \}$  &  $\left \{ \delta_{11},\delta_{21},\delta_{31} \right \}$  \\
		
		$B_{6}$ & & $\left \{ \psi_{1},\psi_{2}  \right \}$  & $\left \{ \delta_{12},\delta_{22},\delta_{32} \right \}$    \\
		
		$B_{7}$ &  & &  $\left \{ \nu_{0},\nu_{1} \right \}$   \\
		
		$B_{8}$ &  &  &   $\left \{ \psi_{1},\psi_{2},\psi_{3}  \right \}$   \\
		
		\hline
		\end{tabular}
\end{table}

The proposal density is a mixture of three multivariate Gaussian proposal distributions, with a random walk mean vector for each block. The proposal variance-covariance matrix of each block in each mixture element is $C_i \Sigma$, where $C_1 =1; C_2 =100; C_3 =0.01$, with $\Sigma$ initially set to $\frac{2.38}{\sqrt{(d_i)}}I_{d_i}$, where $d_{i}$ is the dimension of the $i^{th}$ block and $I_{d_{i}}$ is the identity matrix of dimension $d_{i}$. The vector of mixing weights $(w)$ is (0.7, 0.15, 0.15), allowing both small and large proposal jumps to be considered. The covariance matrix for each block is tuned as in \citet{chen2022dynamic}, with target acceptance rates as in \citet{roberts1997weak}; i.e., acceptance rates: $0.44$ for $d_{i} = 1$, $0.35$ when  $2 \leq d_{i} \leq 4$ and $0.234$ for $d_{i} > 4$. The algorithm is run in epochs, where each epoch is $N = 20,000$ iterations, until the mean total absolute percentage difference in the sample variances of epoch iterates, over all parameters, is less than 10\% (see \citet{chen2022dynamic} for details); typically this takes 3 or 4 epochs. The last 10,000 iterates of the final epoch are used for estimation and inference. 

\section{Data and Empirical study}

\subsection{Data description}

Daily closing prices and RM data from January 2000 to June 2022 were downloaded from Oxford-man Institute’s realized library \citep{heber2009omi}. Three common RMs, including 5-minute Realized Variance (RV5), Realized Kernel (RK), and Bi-power Variation (BV) are considered. Six market indices, including S\&P500 and NASDAQ in the US, FTSE 100 (UK), DAX (Germany), SMI (Swiss), and HSI (Hong Kong), are included in the study. Each data set is split into an initial in-sample period, from January 2000 to December 2011, and an out-of-sample forecasting period from January 2012 to June 2022. Our out-of-sample period includes the COVID-19 period. Figure \ref{Figure 1} displays a time series plot of the absolute value of daily return, RV5, RK and BV of S\&P500 for exposition.
\begin{figure}[H]
\centering
\includegraphics[width=0.7\textwidth]{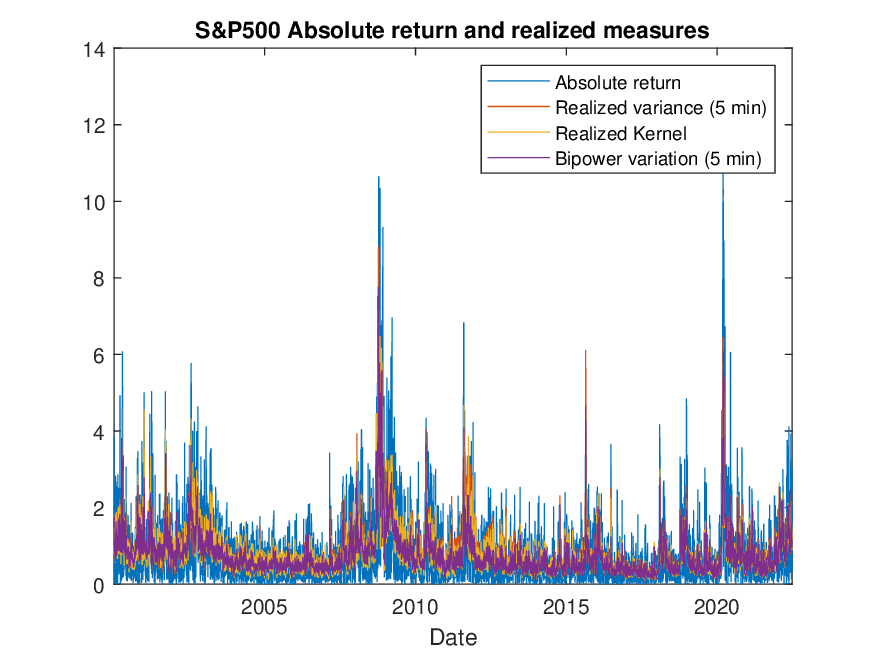}
\caption{S\&P500 absolute return series and three realized measures from January 2000 to June 2022.}
\label{Figure 1}
\end{figure}

Daily one-step-ahead forecasts of VaR and ES are calculated for the six return series at the $1\%$ and $2.5\%$ probability level in the forecast sample. A rolling window, with fixed in-sample size $T$, is used to estimate each of $m$ one-step-ahead forecasts of VaR and ES in the forecast period for each series. Table \ref{table:2} shows the total sample sizes, plus $T$ and $m$ in each market. $T$ and $m$ differ due to different non-trading days in each market.  
\begin{table}[H] 
\caption{Summary of the selected data sets and their in-sample and out-of-sample split.}  \label{table:2}
\small
\centering{}%
\begin{tabular}{lccc}
\hline 
Index & Sample size & In-sample size ($T$) & Out-of-sample size ($m$) \tabularnewline
\hline 
S\&P500 & 5634 & 3008 & 2626\tabularnewline
FTSE & 5667 & 3020 & 2647 \tabularnewline
NASDAQ & 5636 & 3006 & 2630\tabularnewline
HSI & 5504 & 2937 & 2567\tabularnewline
DAX & 5697 & 3050 & 2647\tabularnewline
SMI & 5634 & 3013 & 2621\tabularnewline
\hline 
\end{tabular}
\end{table}

\subsection{Models in comparison}

Table \ref{table:3} lists the 33 models considered in the tail risk forecasting study. As in Section \ref{background_models_section}, four groups of models 
are included, for comparison, as now discussed.

\begin{table}[H]
\centering
\footnotesize
\renewcommand{\arraystretch}{0.4}
	\caption{A summary of the competing models in the empirical section, based on the model type, realized measures used, and the number of realized measures ($K$).}
	\label{table:3}
	\centering
\begin{tabular}{lccc}
\hline
\textbf{Model}        & \textbf{Type}   & \textbf{Realized measures} & \textbf{$K$} \\ 
\rule{0pt}{15pt}\textbf{GARCH Models} & & &\\
\hline 
GARCH-t               & Parametric      & NA                                  & 0                                \\
EGARCH-t              & Parametric      & NA                                  & 0                                \\ 
GJR-GARCH-t           & Parametric      & NA                                  & 0                                \\ 
GARCH-QML-HS             & Parametric      & NA                                  & 0                                \\ 
EGARCH-QML-HS            & Parametric      & NA                                  & 0                                \\ 
GJR-GARCH-QML-HS         & Parametric      & NA                                  & 0                                \\

\rule{0pt}{15pt}\textbf{REGARCH-t Models} & & & \\
\hline 
RV5        & Parametric      & RV5                                 & 1                                \\ 
RK         & Parametric      & RK                                  & 1                                \\ 
BV         & Parametric      & BV                                  & 1                                \\ 
RV5-RK     & Parametric      & RV5, RK                             & 2                                \\ 
RV5-BV     & Parametric      & RV5, BV                             & 2                                \\ 
RK-BV      & Parametric      & RK, BV                              & 2                                \\ 
RV5-RK-BV  & Parametric      & RV5, RK, BV                         & 3                               \\  

\rule{0pt}{15pt}\textbf{ES-CAViaR Models} & & & \\
\hline 
ES-CAViaR-Add         & Semi-Parametric      & NA                                  & 0                                \\  
\rule{0pt}{15pt}\textbf{ES-CAViaR-X Models} & & &\\
\hline 
RV5     & Semi-parametric & RV5   & 1  \\ 
RK      & Semi-parametric & RK   & 1   \\ 
BV      & Semi-parametric & BV   & 1  \\ 

\rule{0pt}{15pt}\textbf{ES-X-CAViaR-X Models} & & &\\
\hline 
RV5 & Semi-parametric & RV5   & 1   \\ 
RK   & Semi-parametric & RK   & 1  \\ 
BV   & Semi-parametric & BV   & 1   \\ 

\rule{0pt}{15pt}\textbf{Realized-ES-CAViaR Models} & & &\\
\hline 
RV5     & Semi-parametric & RV5  & 1   \\
RK      & Semi-parametric & RK  & 1  \\ 
BV      & Semi-parametric & BV   & 1  \\ 

\rule{0pt}{15pt}\textbf{Log-Realized-ES-CAViaR Models} & & &\\
\hline 
RV5     & Semi-parametric & RV5  & 1   \\
RK      & Semi-parametric & RK  & 1  \\ 
BV      & Semi-parametric & BV   & 1  \\ 

\rule{0pt}{15pt}\textbf{Realized-ES-CAViaR-M Models} & & &\\
\hline 
\cellcolor[HTML]{C0C0C0}RV5                & \cellcolor[HTML]{C0C0C0}Semi-parametric & \cellcolor[HTML]{C0C0C0}RV5                                 & \cellcolor[HTML]{C0C0C0}1                                \\ 
\cellcolor[HTML]{C0C0C0}RK                 & \cellcolor[HTML]{C0C0C0}Semi-parametric & \cellcolor[HTML]{C0C0C0}RK                                  & \cellcolor[HTML]{C0C0C0}1                                \\ 
\cellcolor[HTML]{C0C0C0}BV                 & \cellcolor[HTML]{C0C0C0}Semi-parametric &\cellcolor[HTML]{C0C0C0}BV                                  & \cellcolor[HTML]{C0C0C0}1                                \\ 
\cellcolor[HTML]{C0C0C0}RV5-RK             & \cellcolor[HTML]{C0C0C0}Semi-parametric & \cellcolor[HTML]{C0C0C0}RV5, RK                             & \cellcolor[HTML]{C0C0C0}2                                \\ 
\cellcolor[HTML]{C0C0C0}RV5-BV             & \cellcolor[HTML]{C0C0C0}Semi-parametric & \cellcolor[HTML]{C0C0C0}RV5, BV                             & \cellcolor[HTML]{C0C0C0}2                                \\ 
\cellcolor[HTML]{C0C0C0}RK-BV              & \cellcolor[HTML]{C0C0C0}Semi-parametric & \cellcolor[HTML]{C0C0C0}RK, BV                              & \cellcolor[HTML]{C0C0C0}2                                \\ 
\cellcolor[HTML]{C0C0C0}RV5-RK-BV          & \cellcolor[HTML]{C0C0C0}Semi-parametric & \cellcolor[HTML]{C0C0C0}RV5, RK, BV                         & \cellcolor[HTML]{C0C0C0}3                               \\ \hline
\end{tabular}
\begin{flushleft}
    {Note: ``NA'' represents that the model does not use realized measures. Grey shading highlights the proposed models.}
\end{flushleft}
\end{table}

Conventional GARCH \citep{bollerslev1986generalized}, EGARCH \citep{nelson1991conditional}, and GJR-GARCH models \citep{glosten1993relation}, all with Student's $t$ return error are included. The two-step QML-HS approach
as described in Section \ref{garch_section} is also considered, with GARCH, EGARCH and GJR-GARCH employed as the volatility models. 

Next, the parametric REGARCH model with RV5, RK, and BV is included. Student's $t$ return error (REGARCH-t) with Gaussian measurement error is considered, as in \citet{watanabe12}. A similar MCMC algorithm is employed for the estimation of this model. There are seven different versions, which are models with one, two, or three RMs. We have also tested models with Gaussian errors for the return equations, however, these are outperformed by the Student's $t$ return error models and hence not included to save space. 

From the semi-parametric models, ES-CAViaR-X (additive), ES-X-CAViaR-X \citep{gerlach2020semi} and (Log-)Realized-ES-CAViaR ($K=1$) \citep{wang2023semi} are included and estimated via similar adaptive MCMC algorithms. Finally, seven versions of the proposed Realized-ES-CAViaR-M framework are included, again being models with one, two or three RMs included.  

\subsection{Parameter estimates}\label{parameter_estimates_section}
\par

In this section, we study the parameter estimates from the proposed model on 1\% \& 2.5\% probability levels and their comparison to the REGARCH-t, based on parameter inference results for one forecasting step and some discussions for $\boldsymbol{\gamma}$ parameters of the the full out-of-sample period. Only the common parameters between Realized-ES-CAViaR-M and REGARCH-t are compared, since the REGARCH-t does not have the ES component related parameters, e.g., the ones in equation (\ref{eq: 6}).

\subsubsection{One forecasting step results}

For the proposed Realized-ES-CAViaR-M on 1\% \& 2.5\% probability levels and REGARCH-t with 3 RMs, Table \ref{parameter_estimates} shows the parameter posterior means and the lower and upper quantiles (LQ and UQ) of the 95\% credible intervals (CI), using the first moving window of S\&P 500 data. Insignificant parameter estimates with CIs include 0 are highlighted in red. We have the following observations.

First, we find our REGARCH-t parameter estimates are in general consistent with the ones in Tables 2 and 3 of \cite{hansen2016exponential}. The volatility autoregressive parameter $\beta$ estimate is close 1. The $\tau_{1}$ and $\tau_{2}$ estimates are both significant with the value of $\tau_{1}$ as negative, so that negative returns have more impact on the future volatility. Regarding the coefficients $\gamma_{1}$ (coefficient of RV5), $\gamma_{2}$ (RK) and $\gamma_{3}$ (BV) that are used to model the information from RMs, interestingly we see that the $\gamma_{1}$ estimate is insignificant. For the parameters in the measurement equation, $\varphi_{1}$, $\varphi_{2}$ and $\varphi_{3}$ (regression coefficients between RMs and volatility) are all close to unity. Negative $\xi_{1}$, $\xi_{2}$ and $\xi_{3}$ estimates are produced meaning negative bias corrections are needed when regressing RMs versus volatility. This is to be expected, as RMs are only measured when the market is open, while the returns employed in this paper are close-to-close and include overnight price movements, thus downward biased correction is needed. The leverage effect captured in the measurement equation is also significant, with the values of $\delta_{11}$, $\delta_{21}$ and $\delta_{31}$ as negative.

Second, regarding the parameter estimates of the proposed Realized-ES-CAViaR-M model, we observe that the ranges of CIs are in general consistent with the ones from the REGARCH-t. When checking the values of the parameter estimates, we observe some distinctive while explainable behaviours. The $\beta$ estimate is also close unity. However, we can see that the $\tau_{1}$ estimates from the proposed Realized-ES-CAViaR-M are positive. This is because the left tail quantile $Q_t$ has negative values, thus the defined multiplicative error $ \epsilon_{t} = \frac{r_t} {Q_t} $ used in the leverage term have an opposite sign to the return $r_t$ and $z_t$ in REGARCH-t. With respect to the $\gamma_{1}$ (coefficient of RV5), $\gamma_{2}$ (RK) and $\gamma_{3}$ (BV) parameter estimates, it is very interesting to see that on 1\% and 2.5\% different RMs are significant, meaning for different probability levels and potentially for different forecasting steps different RMs could play more important role in risk forecasting. In the full-of-sample study to be shown in the following section, we will have more discussion on this.

Third, for the measurement equation of the proposed Realized-ES-CAViaR-M, although we use $Q_t$ as regressor, we still see that the $\varphi_{1}$, $\varphi_{2}$ and $\varphi_{3}$ estimates are close to unity, which is consistent with the REGARCH-t. Meanwhile, when comparing to REGARCH-t, more negative $\xi_{1}$, $\xi_{2}$ and $\xi_{3}$ estimates are produced. As discussed in Section \ref{proposed_model_section} when developing the proposed model, we have $\sigma_t= \frac{Q_t} {a_{\alpha}}$ with ${a_{\alpha}}$ as a number that is negative with absolute value that is greater than 1 for the considered 1\% and 2.5\% probability levels. For example with a standard Gaussian return distribution, the ${a_{\alpha}}$ values on 1\% and 2.5\% are equal to the Gaussian CDF inverse values of -2.3263 and -1.96, which is also why the 1\% estimated Realized-ES-CAViaR-M has $\boldsymbol{\xi}$ estimates that are more negative than the ones of 2.5\%, e.g., -1.2498 vs -1.0468 for $\xi_1$. Therefore, with more negative bias correction observed, the proposed model is still able to produce close to unity $\boldsymbol{\varphi}$ estimates as REGARCH-t does. Lastly, the leverage term related coefficients are also all significant, while the signs $\delta_{11}$, $\delta_{21}$ and $\delta_{31}$ are again opposite to the ones from REGARCH-t.

\begin{table}[!h]
\begin{center}\footnotesize
\caption{\label{parameter_estimates} \small Parameter posterior means and lower and upper quantiles of the 95\% credible intervals of Realized-ES-CAViaR-M (1\% \& 2.5\%) and REGARCH-t, with the 1st set of in-sample data of S\&P500.}\tabcolsep=5pt
\begin{tabular}{lccc|ccc|ccc} \hline
   &        \multicolumn{3}{c}{Realized-ES-CAViaR-M (1\% )} & \multicolumn{3}{c}{Realized-ES-CAViaR-M (2.5\% )}  &  \multicolumn{3}{c}{REGARCH-t}    \\
                          & Mean    & LB     & UB                         & Mean    & LB     & UB    & Mean    & LB     & UB  \\ \hline
$\omega$&0.0102&0.0040&0.0161&\cred{0.0046}&-0.0007&0.0100&\cred{0.0010}&-0.0080&0.0089\\
$\beta$&0.9717&0.9664&0.9772&0.9705&0.9650&0.9759&0.9660&0.9588&0.9729\\
$\tau_{1}$&0.1812&0.1649&0.1980&0.1526&0.1377&0.1696&-0.1517&-0.1682&-0.1361\\
$\tau_{2}$&0.1160&0.0904&0.1452&0.0850&0.0671&0.1029&0.0464&0.0372&0.0555\\
$\gamma_{1}$&\cred{0.0108}&-0.0318&0.0509&\cred{-0.0046}&-0.0472&0.0386&\cred{-0.0448}&-0.0904&0.0042\\
$\gamma_{2}$&\cred{0.0015}&-0.0257&0.0290&0.0300&0.0031&0.0572&0.0598&0.0303&0.0901\\
$\gamma_{3} $&0.2278&0.1927&0.2649&0.2081&0.1693&0.2486&0.2507&0.2023&0.2978\\
$\xi_{1}$&-1.2498&-1.2899&-1.2097&-1.0468&-1.0882&-1.0046&-0.4296&-0.4889&-0.3706\\
$\xi_{2}$&-1.3719&-1.4152&-1.3308&-1.1675&-1.2093&-1.1258&-0.5665&-0.6228&-0.5115\\
$\xi_{3}  $&-1.3419&-1.3823&-1.2999&-1.1392&-1.1820&-1.0962&-0.6371&-0.6979&-0.5730\\
$\varphi_{1}$&1.0428&1.0088&1.0721&1.0400&1.0048&1.0794&1.0322&0.9876&1.0785\\
$\varphi_{2} $&1.0478&1.0120&1.0812&1.0442&1.0072&1.0825&1.0345&0.9899&1.0802\\
$\varphi_{3} $&1.0536&1.0167&1.0863&1.0530&1.0159&1.0929&1.0456&0.9982&1.0930\\
$\delta_{11}$&0.1123&0.0901&0.1333&0.0939&0.0757&0.1132&-0.0861&-0.1039&-0.0683\\
$\delta_{21}$&0.0636&0.0345&0.0924&0.0533&0.0297&0.0786&-0.0443&-0.0686&-0.0174\\
$\delta_{31} $&0.1754&0.1537&0.1955&0.1457&0.1280&0.1640&-0.1362&-0.1537&-0.1192\\
$\delta_{12}$&0.3635&0.3278&0.4010&0.2464&0.2199&0.2734&0.1111&0.0987&0.123\\
$\delta_{22}$&0.6784&0.6280&0.7291&0.4619&0.4250&0.5011&0.2052&0.1864&0.2236\\
$\delta_{32} $&0.2262&0.1947&0.2591&0.1540&0.1301&0.1778&0.0698&0.0591&0.0807\\
$\nu$&&&&&&&10.5214&7.5990&15.2822\\
\hline
\end{tabular}
\end{center}
\emph{Note}:\small  Insignificant parameter estimates are highlighted in red.
\end{table}

\subsubsection{Full out-of-sample results}\label{section:full_oos}

In the REGARCH-t and proposed Realized-ES-CAViaR-M, the regression coefficients $\boldsymbol{\gamma}$ capture how influential the $K$ lagged RMs are on next period volatility or quantile forecast. Meanwhile, a key contribution of the paper is to incorporate the multiple RMs into the quantile (and ES) forecasting. In the previous section we have observed some interesting behaviours for the $\boldsymbol{\gamma}$ parameters. We now further explore how the $\boldsymbol{\gamma}$ estimates behave for the full out-of-sample period. Figure \ref{gamma_plot_fig} displays the full out-of-sample $\gamma_{1}$ (RV5),  $\gamma_{2}$ (RK) and $\gamma_{3}$ (BV) plots for Realized-ES-CAViaR-M (1\% \& 2.5\%) and REGARCH-t with the S\&P500 data. First, we observe that the general pattern of the $\boldsymbol{\gamma}$ estimates from the 1\% \& 2.5\% Realized-ES-CAViaR-M and the REGARCH-t is consistent, with distinctive behaviours observed. This means the information from RMs is used differently in semi-parametric and parametric risk forecasting. Second, for both 1\% \& 2.5\% risk forecasting and volatility forecasting, the BV seems to be the most influential variable. The implication of this observation on the empirical performance will be discussed in the following section. Third, as shown in Table \ref{parameter_estimates}, for the 1st forecasting step of 1\% Realized-ES-CAViaR-M, both the $\gamma_{1}$ and $\gamma_{2}$ parameter estimates are insignificant, while in the latter forecasting steps the insignificant parameters could become significant, e.g., RV5 during the 2019 period. Further, we observe that for different markets and forecasting steps, the significance of the RMs could vary. Such observations demonstrate that REGARCH-t and the proposed Realized-ES-CAViaR-M are capable of selecting RMs for volatility and risk forecasting tasks using a data driven approach. Meanwhile, this naturally brings up the direction for the next step research. Via incorporating a much larger set of RMs and further developed modelling framework, we could investigate which (types) RMs are more important in volatility and risk forecasting and select such RMs using automatic variable selection technique, such as LASSO \citep{tibshirani1996regression}. 

\begin{figure}[H]
\centering
\includegraphics[width=\textwidth]{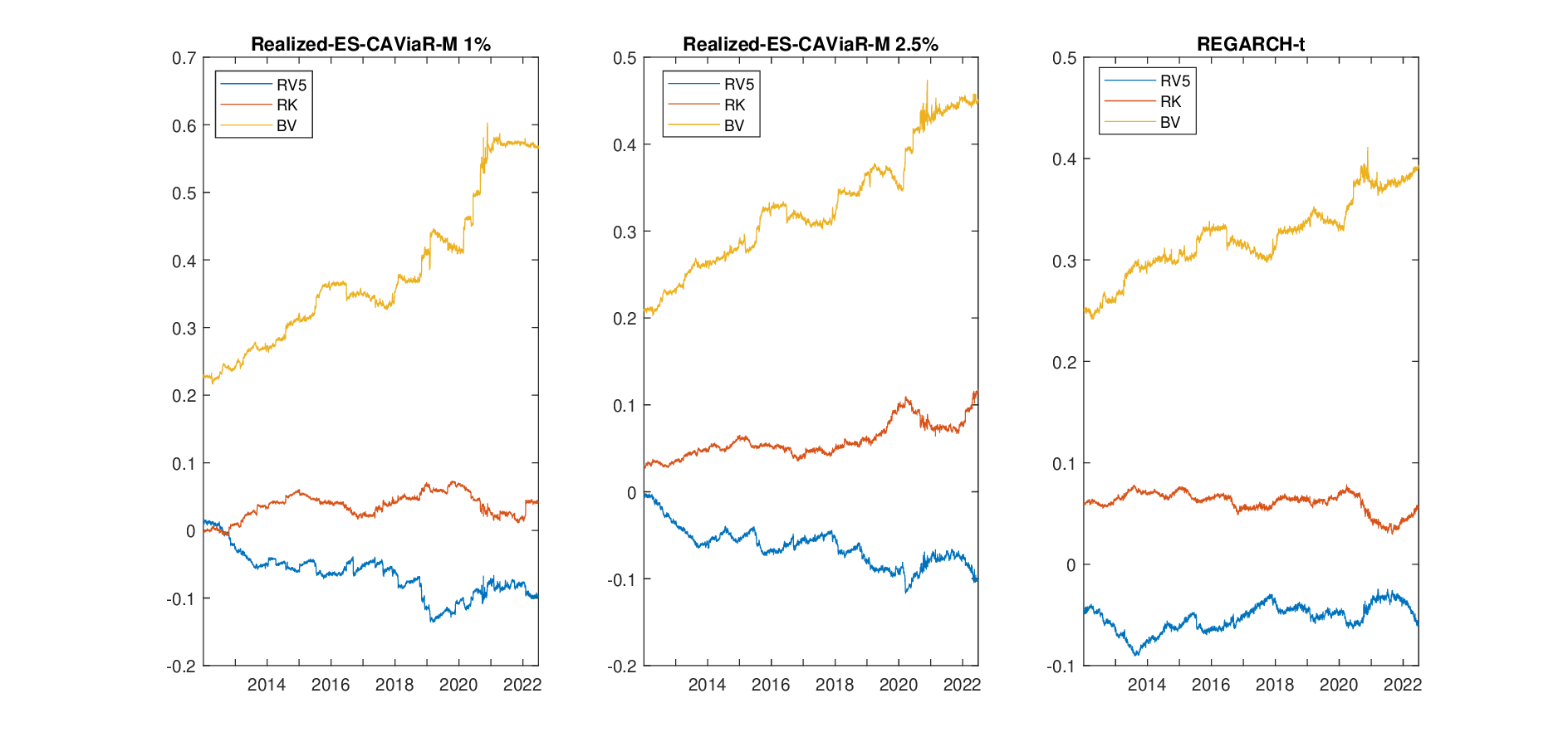}
\caption{The full out-of-sample $\gamma_{1}$ (RV5), $\gamma_{2}$ (RK) and $\gamma_{3}$ (BV) plot for Realized-ES-CAViaR-M (1\% \& 2.5\%) and REGARCH-t with the S\&P500 data.}
\label{gamma_plot_fig}
\end{figure}

\subsection{Assessing Value-at-Risk forecasts}

This section discusses the evaluation of one-step-ahead VaR forecasting accuracy via using the strictly consistent quantile loss function: 
\begin{align}
     \frac{1}{m}\sum_{t = T+1}^{T+m}\left ( \alpha-I(r_{t}\leq \widehat{Q}_{t})\right )\left ( r_{t}-\widehat{Q}_{t}\right ), \label{eq: 14}
\end{align}
where $\widehat{Q}_{T+1},...,\widehat{Q}_{T+m}$ are the quantile forecasts at level $\alpha$. Since the quantile loss function is strictly consistent, the model with minimum sample quantile loss is preferred. Tables \ref{table: q_loss_1} and \ref{table: q_loss_25} present the VaR quantile loss function results on the tested 1\% and 2.5\% probability levels. The average rank based on the ranks of the quantile loss across six markets is also included in the ``Avg Rank'' column. The box indicates the favoured model, and the blue text indicates the second-ranked model in each column.

In general, for both probability levels the proposed Realized-ES-CAViaR-M models produce lower quantile loss values and are better ranked than the other models considered. For the 1\% study, overall the best ranked model is the Realized-ES-CAViaR-M incorporating all three realized measures, followed by the Realized-ES-CAViaR-M with RV5 and BV. For the 2.5\% study, the top two performing models for each market are all from the Realized-ES-CAViaR-M model class. The overall best ranked model is again Realized-ES-CAViaR with three realized measures, followed by Realized-ES-CAViaR-M with BV. The preferred performance of Realized-ES-CAViaR-M with three RMs demonstrate the effectiveness of incorporating multiple RMs in semi-parametric quantile forecasting. Further, based on the parameter estimates in Section \ref{parameter_estimates_section}, BV seems to be the most influential RM in the Realized-ES-CAViaR-M forecasting process. Such findings are supported by the results that the proposed models with BV are generally preferred and better ranked than the ones without BV. This means that although RK\&RV are potentially less effective than BV in semi-parametric quantile forecasting, the Realized-ES-CAViaR-M with three RMs can still effective incorporate the useful information from RK\&RV and combine it with BV (see Figure \ref{gamma_plot_fig}) in generating improved quantile forecasts.

Compared to the Realized-ES-CAViaR model of \cite{wang2023semi}, the Realized-ES-CAViaR-M extends it via incorporating the information of multiple realized measures during the VaR and ES forecasting process. Further, a log specification is used in Realized-ES-CAViaR-M, with an additional leverage employed in the quantile equation (\ref{eq: 5}). Based on the empirical results, now we investigate the improved performance of the Realized-ES-CAViaR-M models is due to the log specification, the leverage term, or the inclusion of multiple RMs. 
    
Regarding the log-specification, comparing the Realized-ES-CAViaR and Log-Realized-ES-CAViaR, the overall consistent performance of the two classes of models shows that the log specification does not significantly affect the risk forecasting performance.

Regarding the leverage effect, actually the (Log-)Realized-ES-CAViaR model of \cite{wang2023semi} has already considered it in its quantile regression process. For example, in the Log-Realized-ES-CAViaR model (\ref{Log_Realized_ES_CAViaR_model}) by substituting the $\mathrm{log}(x_{t-1})$ in the quantile equation with its expression in the measurement equation we get: 
$$\mathrm{log}(-Q_{t})=\beta _{0}+\beta _{1} \left( \xi+\phi \mathrm{log}(-Q_{t-1})+\tau_{1}\epsilon_{t-1}+ \tau_{2}\left ( \epsilon^2_{t-1}-E(\epsilon^2) \right ) +u_{t-1} \right)+\beta _{2} \mathrm{log} (-Q_{t-1}),$$
thus the leverage effect is considered in its VaR forecasting. Consequently, as ES is equal to VaR subtracting $\omega_t$, the leverage effect is also implicitly considered in the ES forecasting process.
    
Therefore, the overall favoured performance of the Realized-ES-CAViaR-M with three RMs confirms the key driver of its improved risk forecasting performance is the information contained in the multiple RMs, which is captured by the proposed framework.

The performance of the REGARCH-t on the 2.5\% level is slightly better than the more extreme 1\% level, demonstrating that fixing the return distribution for different probability levels could limit the performance of the parametric model. Meanwhile, the superior performance of Realized-ES-CAViaR-M shows the advantages of semi-parametric risk forecasting, as the selection of return error distribution is not required.

To statistically test whether the quantile loss differences between different models are significant, the model confidence set (MCS), introduced by \citet{hansen2011model}, is employed. The MCS produces a group of models that is constructed such that it will contain the ``superior'' forecasting models, given a level of confidence. The MCS is used to assess the statistical significance for quantile loss (per equation (\ref{eq: 14})) under the 75\% confidence level. We adopt the Matlab code downloaded from Kevin Sheppard's web page \citep{sheppard2009mfe}. Two methods, R and SQ, are calculated to test the competing models based on different rules of calculating the test statistic in the downloaded MCS code. In Tables \ref{table: q_loss_1} and \ref{table: q_loss_25}, for each market we use grey shading to highlight the models included in MCS, based on the R method, for visual comparison of the models. As can be seen, the proposed Realized-ES-CAViaR-M models are more or equally likely to be included in the MCS, compared to other models. The GARCH-type models and ES-CAViaR-Add (without using realized data) are generally less likely to be included in the MCS, demonstrating the usefulness of incorporating realized data in either parametric or semi-parametric risk forecasting. In addition, Table \ref{table: MCS} presents a summary of MCS results based on the quantile loss for both R and SQ methods. The R and SQ columns show the total number of times that each model is included in the $75\%$ MCS across the six return series (the higher, the better). The results from the SQ method are consistent with the ones from the R method, with the Realized-ES-CAViaR-M type models being more or equally likely to be included in MCS compared to other models. The Realized-ES-CAViaR-M with three RMs is the only model that is always in the MCS across six markets for both tests and probability levels.

\subsection{Assessing Expected Shortfall forecasts}

The same 33 models are employed to generate one-step-ahead ES forecasts at the 1\% and 2.5\% probability levels for the same six series in the forecasting period. 

As discussed in Section \ref{likelihood_function}, \citet{taylor2019forecasting} shows that the negative of the quasi-log-likelihood function (\ref{al_likelihood}) is strictly consistent for $Q_t$ and $\text{ES}_t$ considered jointly, and fits into the class of strictly consistent joint loss functions for VaR and ES developed by \citet{fissler2016higher}. We use the average joint loss $S = \frac{1}{m} \sum_{t=n+1}^{n+m} S_t$ to formally and jointly assess the VaR and ES forecasts from all models.
\begin{eqnarray}\label{es_caviar_log_score}
S_t(r_t, \widehat{Q}_t, \widehat{\text{ES}}_t) = -\text{log} \left( \frac{\alpha-1}{\widehat{\text{ES}}_t} \right) - {\frac{(r_t- \widehat{Q}_t)(\alpha-I(r_t\leq \widehat{Q}_t))}{\alpha \widehat{\text{ES}}_t}}.
\end{eqnarray}

Tables \ref{table: joint_1} and \ref{table: joint_25} show the 1\% and 2.5\% VaR and ES joint loss function values for each model and each market. The ``Avg Rank'' column calculated under the same way as Tables \ref{table: q_loss_1} and \ref{table: q_loss_25} is included. Grey shading is again used to highlight the models included in the 75\% MCS, based on the R method with joint loss. The results are generally consistent with the ones from the quantile loss study. For each market, on the 1\% level the top ranked models are more likely from the proposed model class, and on the 2.5\% level all the top 2 ranked models are from the proposed Realized-ES-CAViaR-M models. Again, the overall best ranked model is the Realized-ES-CAViaR-M with all three realized measures. Although the parametric REGARCH-t with three RMs could generate competitive forecasting results, it is still consistently outperformed by its semi-parametric counter-part, the proposed Realized-ES-CAViaR-M, again demonstrating the usefulness of forecasting the risk without assuming the return error distribution. The performance of the Realized-ES-CAViaR and Log-Realized-ES-CAViaR are similar to each other, and they are outperformed by the Realized-ES-CAViaR-M with three realized measures. Such observations demonstrate the effectiveness of employing the information from multiple RMs in semi-parametric ES forecasting. The MCS results in Tables \ref{table: joint_1} and \ref{table: joint_25} support the Realized-ES-CAViaR-M model as being more or equally likely to be included in the MCS. A summary of the joint loss-based MCS results based on R and SQ methods are presented in Table \ref{table: MCS}. The results from the joint loss SQ method align with the ones from the R method. Across six markets, the Realized-ES-CAViaR-M with three RMs is the only model that is always in the MCS for both tests and probability levels.

Another interesting observation is that when the same realized measure(s) are included in the Realized-ES-CAViaR-M models, their VaR forecasting performance based on quantile loss could differ to the one from VaR\&ES joint loss. For example, via comparing the rank results in Tables \ref{table: q_loss_25} and \ref{table: joint_25} on the 2.5\% probability level, when only one realized measure is included in the Realized-ES-CAViaR-M models, they performed more competitively in terms of quantile loss than joint loss. Such observation is potentially related to the additive VaR to ES time varying relationship as described in equations (\ref{eq: 6}) and (\ref{eq: 7}), driven \emph{separately} by the lagged RMs comparing to the ones in the quantile equation (\ref{eq: 5}). There is potentially interesting future work based on such observation:  which (types of) realized measures could be more useful in VaR forecasting? Is it the case that other (types of) realized measures more useful in ES forecasting? These questions could be investigated via studying a large set of RMs in VaR and ES forecasting via a further developed modelling framework. The LASSO approach can be used to conduct automatic variable selection, as discussed at the end of Section \ref{section:full_oos}.

To summarize, with the 1\% and 2.5\% quantile and joint loss evaluations and MCS backtests, for VaR and ES forecasting accuracy comparison over six markets, the proposed Realized-ES-CAViaR-M framework has generally favourable performance compared to a range of competing models. The performance is most favourable for the proposed model using all three RMs.

\setlength{\tabcolsep}{2pt}
\begin{table}[H]    
\centering
\footnotesize
\renewcommand{\arraystretch}{0.4}
	\caption{$1\%$ VaR forecasting quantile loss on six indices.}
	\label{table: q_loss_1}
	\centering
	\begin{tabular}{p{1.8 in}cccccc|c}
		\hline
	\textbf{Model}  & \textbf{S\&P500} & \textbf{FTSE} & \textbf{NASDAQ}  & \textbf{HSI}  & \textbf{DAX} & \textbf{SMI}   & \textbf{Avg Rank} \\
 \hline
         \rule{0pt}{15pt}\textbf{GARCH} &&&&&&& \\
        \hline 
        
GARCH-t   &89.8&91.4&105.8&\cg{100.4}&\cg{111.1}&88.0&32.7\\
EGARCH-t &87.5&88.1&102.6&\cg{96.2}&\cg{107.5}&85.5&23.8\\
GJR-GARCH-t &89.0&\cg{86.8}&103.2&\cg{96.4}&\cg{108.6}&\cg{84.8}&25.0\\
GARCH-QML-HS&88.3&88.1&101.0&\cg{98.0}&\cg{110.3}&86.5&29.5\\
EGARCH-QML-HS &86.5&\cg{86.1}&\cg{99.0}&\cg{95.4}&\cg{\fb{106.4}}&\cg{84.4}&17.5\\
GJR-GARCH-QML-HS&87.9&\cg{84.3}&100.9&\cg{95.4}&\cg{107.5}&\cg{84.1}&17.7\\
           \rule{0pt}{15pt}\textbf{REGARCH-t} &&&&&&& \\
        \hline 
RV5&85.2&88.7&\cg{97.1}&97.9&\cg{110.1}&84.5&26.3\\
RK &80.9&88.1&\cg{95.1}&\cg{97.2}&\cg{114.8}&86.1&25.5\\
BV &82.7&88.9&\cg{95.5}&100.4&\cg{108.9}&86.1&25.0\\
RV5-RK  &83.5&87.6&\cg{94.1}&\cg{95.1}&\cg{109.9}&\cg{82.6}&18.7\\
RV5-BV  &80.6&87.9&\cg{93.8}&100.3&\cg{108.8}&\cg{82.6}&19.2\\
RK-BV  &80.1&87.9&\cg{93.8}&\cg{95.5}&\cg{109.1}&\cg{83.3}&16.8\\
RV5-RK-BV &79.7&87.3&\cg{93.9}&\cg{95.5}&\cg{108.9}&\cg{82.6}&15.5\\
	  \rule{0pt}{15pt}\textbf{ES-CAViaR} &&&&&&& \\
        \hline 
ES-CAViaR-Add&87.2&89.3&100.4&\cg{97.7}&\cg{112.4}&87.1&29.5\\
	  \rule{0pt}{15pt}\textbf{ES-CAViaR-X} &&&&&&& \\
        \hline 
RV5&87.8&\cg{85.3}&\cg{93.0}&\cg{95.7}&\cg{109.9}&\cg{82.2}&17.2\\
RK &87.3&\cg{83.2}&\cg{93.2}&\cg{96.0}&\cg{109.6}&84.5&18.3\\
BV &88.2&\cg{84.4}&\cg{91.9}&\cg{95.9}&\cg{107.7}&\cg{\fb{80.6}}&12.2\\
	  \rule{0pt}{15pt}\textbf{ES-X-CAViaR-X} &&&&&&& \\
        \hline 
RV5&82.8&\cg{84.6}&\cg{93.0}&\cg{95.5}&\cg{110.0}&\cg{82.3}&15.7\\
RK &82.4&\cg{83.4}&\cg{93.9}&\cg{95.9}&\cg{110.5}&84.6&19.2\\
BV &79.2&\cg{84.4}&\cg{92.1}&\cg{96.0}&\cg{108.3}&\cg{\cb{80.8}}&9.8\\
        \rule{0pt}{15pt}\textbf{Realized-ES-CAViaR} &&&&&&& \\
        \hline 
RV5&83.1&\cg{85.1}&\cg{93.2}&\cg{95.0}&\cg{110.1}&\cg{82.1}&14.8\\
RK &84.0&\cg{85.5}&\cg{94.2}&\cg{95.5}&\cg{111.0}&\cg{83.9}&21.3\\
BV &\cg{78.5}&\cg{85.6}&\cg{92.1}&\cg{95.8}&\cg{108.9}&\cg{82.1}&12.2\\
\rule{0pt}{15pt}\textbf{Log Realized-ES-CAViaR} &&&&&&& \\
        \hline 
RV5&82.2&\cg{85.2}&\cg{92.5}&\cg{95.0}&\cg{109.9}&\cg{82.2}&13.0\\
RK &83.1&\cg{84.4}&\cg{93.6}&\cg{96.2}&\cg{110.3}&\cg{84.0}&19.8\\
BV &\cg{78.6}&\cg{84.9}&\cg{\fb{90.8}}&\cg{95.8}&\cg{109.1}&\cg{81.6}&10.5\\
        \rule{0pt}{15pt}\textbf{Realized-ES-CAViaR-M} &&&&&&& \\
        \hline 
RV5&80.9&\cg{85.4}&\cg{92.7}&\cg{\cb{94.4}}&\cg{108.2}&\cg{81.4}&9.2\\
RK &81.7&\cg{83.0}&\cg{93.7}&\cg{97.3}&\cg{107.9}&\cg{82.9}&14.5\\
BV &\cg{78.0}&\cg{83.6}&\cg{92.0}&\cg{95.2}&\cg{107.1}&\cg{82.3}&7.2\\
RV5-RK  &82.6&\cg{83.5}&\cg{92.8}&\cg{\fb{94.3}}&\cg{110.0}&\cg{81.6}&10.8\\
RV5-BV  &\cg{\cb{77.9}}&\cg{83.6}&\cg{92.1}&\cg{95.0}&\cg{106.5}&\cg{81.5}&\cb{4.8}\\
RK-BV  &\cg{78.0}&\cg{\cb{82.9}}&\cg{\cb{91.7}}&\cg{95.7}&\cg{\cb{106.6}}&\cg{81.4}&5.2\\
RV5-RK-BV &\cg{\fb{77.8}}&\cg{\fb{82.8}}&\cg{91.9}&\cg{94.8}&\cg{107.0}&\cg{81.3}&\fb{2.7}\\
		\hline
	\end{tabular}
 \begin{flushleft}
    {Note: The box indicates the favoured models, and the blue text indicates the second-ranked model in each column. Grey shades the models that are included in the 75\% MCS using the R method.}
\end{flushleft}
\end{table}

\setlength{\tabcolsep}{2pt}
\begin{table}[H]    
\centering
\footnotesize
\renewcommand{\arraystretch}{0.4}
	\caption{$2.5\%$ VaR forecasting quantile loss on six indices.}
	\label{table: q_loss_25}
	\centering
	\begin{tabular}{p{1.8 in}cccccc|c}
		\hline
	\textbf{Model}  & \textbf{S\&P500} & \textbf{FTSE} & \textbf{NASDAQ}  & \textbf{HSI}  & \textbf{DAX} & \textbf{SMI}   & \textbf{Avg Rank} \\
		
        \rule{0pt}{15pt}\textbf{GARCH} &&&&&&& \\
        \hline 
		GARCH-t   & 181.2  &  183.0 & 218.5   &    210.9& 226.0 & 175.0  & 31.8     \\
		
		EGARCH-t & 175.3 &  \cellcolor[HTML]{C0C0C0}176.9 & 213.4  & \cellcolor[HTML]{C0C0C0}203.7 & \cellcolor[HTML]{C0C0C0}217.2 &  168.6   & 20.5     	 \\
		GJR-GARCH-t & 174.6 & \cellcolor[HTML]{C0C0C0}176.8 &  211.1 & \cellcolor[HTML]{C0C0C0}204.7 &  \cellcolor[HTML]{C0C0C0}219.9 &   168.0  & 21.2     \\
		
		GARCH-QML-HS& 182.0 &  184.0 & 219.8  &   211.9 & 227.3 & 175.4 & 32.8 \\
		
		EGARCH-QML-HS & 176.7 &  \cellcolor[HTML]{C0C0C0}177.9 & 214.8  & \cellcolor[HTML]{C0C0C0}204.8 & \cellcolor[HTML]{C0C0C0}217.6  &  169.0  & 23.0 \\
		
		GJR-GARCH-QML-HS& 175.5 & \cellcolor[HTML]{C0C0C0}177.6 &  212.3  & \cellcolor[HTML]{C0C0C0}205.4 & \cellcolor[HTML]{C0C0C0}220.3  &  168.4  & 24.3\\

           \rule{0pt}{15pt}\textbf{REGARCH-t} &&&&&&& \\
        \hline 
        RV5  &  183.7 &  177.9 & \cellcolor[HTML]{C0C0C0}203.3 &  204.2 & 223.1 &  166.9  & 24.7\\
		
		RK & \cellcolor[HTML]{C0C0C0}169.6 &  177.9 & \cellcolor[HTML]{C0C0C0}201.1 & \cellcolor[HTML]{C0C0C0}202.7 & 223.9 & 168.7  & 21.8 \\
		
		BV & 172.8 & \cellcolor[HTML]{C0C0C0}178.3 & \cellcolor[HTML]{C0C0C0}201.4 & \cellcolor[HTML]{C0C0C0}205.3 & 221.9 &  170.6  & 25.2 \\
				
		RV5-RK  & \cellcolor[HTML]{C0C0C0}171.4 & \cellcolor[HTML]{C0C0C0}176.5 &  \cellcolor[HTML]{C0C0C0}197.8 & \cellcolor[HTML]{C0C0C0}199.9 & 220.6 & \cellcolor[HTML]{C0C0C0}164.7  & 13.7 \\
				
		RV5-BV  & \cellcolor[HTML]{C0C0C0}168.0 &  177.8 & \cellcolor[HTML]{C0C0C0}197.9 & 205.7 & 220.7 & \cellcolor[HTML]{C0C0C0}164.7  & 18.5 \\
				
		RK-BV  & \cellcolor[HTML]{C0C0C0}168.0 & \cellcolor[HTML]{C0C0C0}176.9 &  \cellcolor[HTML]{C0C0C0}197.0 & \cellcolor[HTML]{C0C0C0}199.5& 220.0 & \cellcolor[HTML]{C0C0C0}165.5  & 11.8 \\
				
		RV5-RK-BV & \cellcolor[HTML]{C0C0C0}167.8 & \cellcolor[HTML]{C0C0C0}176.4 &  \cellcolor[HTML]{C0C0C0}197.5 & \cellcolor[HTML]{C0C0C0}199.9 & 219.8 & \cellcolor[HTML]{C0C0C0}164.7  & 11.0 \\

  	  \rule{0pt}{15pt}\textbf{ES-CAViaR} &&&&&&& \\
        \hline 
        ES-CAViaR-Add  & 179.2 &  179.7 & 212.0 &  205.4 & 222.9 & 174.6   & 29.0    \\
		
	  \rule{0pt}{15pt}\textbf{ES-CAViaR-X} &&&&&&& \\
        \hline 
        RV5  & 179.0 &  179.6 &   209.9  & \cellcolor[HTML]{C0C0C0}203.1  & 221.6  & 172.5   &  26.3    \\
		
		RK  & 177.6 &  182.0 & \cellcolor[HTML]{C0C0C0}207.3  & \cellcolor[HTML]{C0C0C0}203.6 & 221.5 & 173.3   & 26.5  \\
		
		BV  & 180.5 & \cellcolor[HTML]{C0C0C0}178.9 &  209.3   & \cellcolor[HTML]{C0C0C0}203.1 & 221.3 &  169.9 & 25.3  \\
		
		\rule{0pt}{15pt}\textbf{ES-X-CAViaR-X} &&&&&&& \\
        \hline 
        RV5  & \cellcolor[HTML]{C0C0C0}169.7  & \cellcolor[HTML]{C0C0C0}177.2 & \cellcolor[HTML]{C0C0C0}198.0  & \cellcolor[HTML]{C0C0C0}201.0 & \cellcolor[HTML]{C0C0C0}218.8 & \cellcolor[HTML]{C0C0C0}164.8   & 14.3   \\
		
		RK  & \cellcolor[HTML]{C0C0C0}168.2 & \cellcolor[HTML]{C0C0C0}176.3 &  \cellcolor[HTML]{C0C0C0}197.3  & \cellcolor[HTML]{C0C0C0}202.4 & \cellcolor[HTML]{C0C0C0}218.8 &  166.4   & 13.3   \\
		
		BV  & \cellcolor[HTML]{C0C0C0}168.1 & \cellcolor[HTML]{C0C0C0}175.0 & \cellcolor[HTML]{C0C0C0}196.2   & \cellcolor[HTML]{C0C0C0}200.6 & 219.3 & \cellcolor[HTML]{C0C0C0}166.2  & 11.2  \\
		
        \rule{0pt}{15pt}\textbf{Realized-ES-CAViaR} &&&&&&& \\
        \hline 
       RV5  & \cellcolor[HTML]{C0C0C0}170.7 & \cellcolor[HTML]{C0C0C0}178.3 & \cellcolor[HTML]{C0C0C0}198.0  & \cellcolor[HTML]{C0C0C0}200.6 & \cellcolor[HTML]{C0C0C0}219.0 & 166.6  & 17.2   \\
		
		RK  & \cellcolor[HTML]{C0C0C0}171.3 & \cellcolor[HTML]{C0C0C0}179.0 &  \cellcolor[HTML]{C0C0C0}197.7  & \cellcolor[HTML]{C0C0C0}201.8 & \cellcolor[HTML]{C0C0C0}218.6& 168.0  & 18.3 \\
		
		BV  & \cellcolor[HTML]{C0C0C0}165.8 & \cellcolor[HTML]{C0C0C0}176.8 & \cellcolor[HTML]{C0C0C0}196.1 & \cellcolor[HTML]{C0C0C0}200.9 & \cellcolor[HTML]{C0C0C0}218.2 &  167.9 & 11.0  \\
\rule{0pt}{15pt}\textbf{Log Realized-ES-CAViaR} &&&&&&& \\
        \hline 
       RV5   & \cellcolor[HTML]{C0C0C0}169.9 & \cellcolor[HTML]{C0C0C0}177.1  & \cellcolor[HTML]{C0C0C0}197.3 & \cellcolor[HTML]{C0C0C0}199.9 & \cellcolor[HTML]{C0C0C0}218.3 & \cellcolor[HTML]{C0C0C0}165.7   &  13.0  \\
		
		RK   & \cellcolor[HTML]{C0C0C0}169.8& \cellcolor[HTML]{C0C0C0}176.5 & \cellcolor[HTML]{C0C0C0}196.8 & \cellcolor[HTML]{C0C0C0}205.7 & \cellcolor[HTML]{C0C0C0}218.5 & 167.1  & 13.8   \\
		
		BV   & \cellcolor[HTML]{C0C0C0}165.6& \cellcolor[HTML]{C0C0C0}174.6 & \cellcolor[HTML]{C0C0C0}195.8 & \cellcolor[HTML]{C0C0C0}199.5 & \cellcolor[HTML]{C0C0C0}218.5 & \cellcolor[HTML]{C0C0C0}166.6   &  9.5  \\

        \rule{0pt}{15pt}\textbf{Realized-ES-CAViaR-M} &&&&&&& \\
        \hline 
		RV5  & \cellcolor[HTML]{C0C0C0}166.1 & \cellcolor[HTML]{C0C0C0}173.7 & \cellcolor[HTML]{C0C0C0}\color[HTML]{3166FF}194.8  & \cellcolor[HTML]{C0C0C0}\color[HTML]{3166FF}197.9 &\cellcolor[HTML]{C0C0C0}216.7 &   \cellcolor[HTML]{C0C0C0}\color[HTML]{3166FF}162.9   & 3.8   \\
		RK  & \cellcolor[HTML]{C0C0C0}166.3 & \cellcolor[HTML]{C0C0C0}173.4 & \cellcolor[HTML]{C0C0C0}195.5  &  \cellcolor[HTML]{C0C0C0}\framebox[0.08\columnwidth]{197.7} & \cellcolor[HTML]{C0C0C0}215.6 &   \cellcolor[HTML]{C0C0C0}164.2   & 4.2  \\
		BV  &  \cellcolor[HTML]{C0C0C0}\framebox[0.08\columnwidth]{163.2} & \cellcolor[HTML]{C0C0C0}174.1 & \cellcolor[HTML]{C0C0C0}\framebox[0.08\columnwidth]{194.1} &\cellcolor[HTML]{C0C0C0}198.6 & \cellcolor[HTML]{C0C0C0}214.9 &   \cellcolor[HTML]{C0C0C0}164.0   & \color[HTML]{3166FF}3.2    \\
		RV5-RK  & \cellcolor[HTML]{C0C0C0}166.4 & \cellcolor[HTML]{C0C0C0}174.8 &\cellcolor[HTML]{C0C0C0}198.4  &\cellcolor[HTML]{C0C0C0}198.3 & 224.3 &    168.6   & 15.7    \\
		RV5-BV  & \cellcolor[HTML]{C0C0C0}165.3 & \cellcolor[HTML]{C0C0C0}\color[HTML]{3166FF}172.9 & \cellcolor[HTML]{C0C0C0}204.3  &\cellcolor[HTML]{C0C0C0}205.2 & \cellcolor[HTML]{C0C0C0}216.0 &   \cellcolor[HTML]{C0C0C0}165.6   & 12.0    \\
		RK-BV  & \cellcolor[HTML]{C0C0C0}165.1 &  180.2 & \cellcolor[HTML]{C0C0C0}197.4  &\cellcolor[HTML]{C0C0C0}202.2& \cellcolor[HTML]{C0C0C0}\framebox[0.08\columnwidth]{211.7} &   \cellcolor[HTML]{C0C0C0}163.5   & 11.0   \\
	    RV5-RK-BV & \cellcolor[HTML]{C0C0C0}\color[HTML]{3166FF}164.6 & \cellcolor[HTML]{C0C0C0}\framebox[0.08\columnwidth]{172.5} & \cellcolor[HTML]{C0C0C0}195.1  &\cellcolor[HTML]{C0C0C0}198.0 & \cellcolor[HTML]{C0C0C0}\color[HTML]{3166FF}214.7 &    \cellcolor[HTML]{C0C0C0}\framebox[0.08\columnwidth]{162.6}   &  \framebox[0.08\columnwidth]{2.0}   \\

		\hline
	\end{tabular}
 \begin{flushleft}
    {Note: The box indicates the favoured models, and the blue text indicates the second-ranked model in each column. Grey shades the models that are included in the 75\% MCS using the R method.}
\end{flushleft}
\end{table}

\setlength{\tabcolsep}{3pt}
\begin{table}[H]   
\centering
\footnotesize
\renewcommand{\arraystretch}{0.6}
	\caption{$1\%$ VaR \& ES joint loss function values across six indices.}
	\label{table: joint_1}
	\centering
	\begin{tabular}{p{1.8 in}cccccc|c}
		\hline
	\textbf{Model}  & \textbf{S\&P500} & \textbf{FTSE} & \textbf{NASDAQ}  & \textbf{  HSI  }  &  \textbf{ 
 DAX  } & \textbf{  SMI  } & \textbf{Avg Rank}   \\
        \rule{0pt}{15pt}\textbf{GARCH} &&&&&&& \\
        \hline 		
GARCH-t   &5879.8&5881.2&6349.3&\cg{6050.6}&\cg{6365.9}&5681.8&30.0\\
EGARCH-t &5885.0&5928.9&6444.7&\cg{5973.9}&\cg{6366.0}&5686.3&30.5\\
GJR-GARCH-t &5934.1&5823.2&6361.8&\cg{5970.7}&\cg{6390.5}&\cg{5652.1}&30.0\\
GARCH-QML-HS&5726.0&5734.9&6121.3&\cg{5969.0}&\cg{6329.6}&\cg{5588.3}&25.7\\
EGARCH-QML-HS &5717.9&5769.3&6179.2&\cg{5932.7}&\cg{6282.6}&\cg{5573.6}&23.8\\
GJR-GARCH-QML-HS&5749.1&5686.1&6167.0&\cg{5920.7}&\cg{6311.0}&\cg{5570.4}&22.5\\
     \rule{0pt}{15pt}\textbf{REGARCH-t} &&&&&&& \\
        \hline 
RV5&5623.4&5842.3&6059.8&6003.9&6422.3&5587.5&28.3\\
RK &\cg{5364.9}&5794.7&\cg{6053.8}&\cg{5937.2}&6598.9&5645.5&24.8\\
BV &5539.6&5797.1&\cg{5942.7}&\cg{6052.4}&\cg{6380.5}&\cg{5595.8}&26.2\\
RV5-RK  &5511.5&5780.0&\cg{5970.8}&\cg{5917.0}&6449.2&\cg{5504.4}&23.5\\
RV5-BV  &5418.2&5775.0&\cg{5984.3}&6070.6&6393.9&\cg{5484.3}&23.3\\
RK-BV  &5370.1&5768.2&\cg{5966.7}&\cg{5938.2}&6400.5&\cg{5514.2}&22.0\\
RV5-RK-BV &\cg{5350.7}&5734.7&\cg{5960.9}&\cg{5931.8}&6385.2&\cg{5493.9}&19.2\\
        \rule{0pt}{15pt}\textbf{ES-CAViaR} &&&&&&& \\
        \hline 
ES-CAViaR-Add&5692.0&5758.5&6069.2&\cg{5977.1}&\cg{6404.2}&5620.4&27.5\\
        \rule{0pt}{15pt}\textbf{ES-CAViaR-X}  &&&&&&& \\
        \hline 
RV5&5547.1&5597.8&\cg{5802.4}&\cg{5901.2}&\cg{6318.1}&\cg{5416.8}&15.8\\
RK &5568.5&5568.7&\cg{5793.3}&\cg{5896.2}&\cg{6311.2}&\cg{5522.1}&15.2\\
BV &5466.3&5605.9&\cg{5780.6}&\cg{5922.5}&\cg{6257.9}&\cg{\cb{5370.4}}&11.8\\
        \rule{0pt}{15pt}\textbf{ES-X-CAViaR-X} &&&&&&& \\
        \hline 
RV5&5433.5&5582.8&\cg{5800.6}&\cg{5896.1}&\cg{6315.5}&\cg{5407.5}&11.8\\
RK &5453.1&\cg{5566.9}&\cg{5816.2}&\cg{5888.4}&\cg{6315.0}&\cg{5509.2}&14.3\\
BV &\cg{5309.8}&5594.0&\cg{5783.4}&\cg{5926.7}&\cg{6255.8}&\cg{\fb{5363.8}}&9.2\\
        \rule{0pt}{15pt}\textbf{Realized-ES-CAViaR} &&&&&&& \\
        \hline 
RV5&5438.4&5583.8&\cg{5792.4}&\cg{\cb{5874.3}}&\cg{6279.7}&\cg{5432.4}&10.7\\
RK &5479.9&5587.8&\cg{5814.6}&\cg{5876.2}&\cg{6295.7}&\cg{5504.1}&14.2\\
BV &\cg{5293.7}&5593.9&\cg{5779.5}&\cg{5912.5}&\cg{6239.6}&\cg{5414.4}&8.0\\
        \rule{0pt}{15pt}\textbf{Log-Realized-ES-CAViaR} &&&&&&& \\
        \hline 
RV5&5394.5&5595.6&\cg{\cb{5773.8}}&\cg{5881.8}&\cg{6327.9}&\cg{5422.8}&11\\
RK &5433.8&\cg{5567.7}&\cg{5803.9}&\cg{5897.1}&\cg{6314.1}&\cg{5497.8}&13.7\\
BV &\cg{5299.3}&5591.5&\cg{\fb{5752.0}}&\cg{5920.4}&\cg{6272.3}&\cg{5399.3}&8.8\\
        \rule{0pt}{15pt}\textbf{Realized-ES-CAViaR-M} &&&&&&& \\
        \hline 
RV5&5414.6&5708.7&\cg{5806.8}&\cg{5890.6}&\cg{6270.7}&\cg{5406.6}&11.8\\
RK &5445.3&\cg{5564.2}&\cg{5834.3}&\cg{5913.6}&\cg{6258.2}&\cg{5459.4}&13.2\\
BV &\cg{5298.2}&5560.7&\cg{5792.2}&\cg{5914.8}&\cg{6204.8}&\cg{5425.8}&8.3\\
RV5-RK  &5420.6&5567.8&\cg{5800.9}&\cg{\fb{5873.2}}&\cg{6327.3}&\cg{5421.6}&11.2\\
RV5-BV  &\cg{5291.4}&\cg{5535.5}&\cg{5794.6}&\cg{5908.9}&\cg{\cb{6184.6}}&\cg{5401.2}&6.2\\
RK-BV  &\cg{\cb{5289.0}}&\cg{\cb{5522.3}}&\cg{5784.8}&\cg{5914.5}&\cg{\fb{6182.6}}&\cg{5392.8}&5.0\\
RV5-RK-BV &\cg{\fb{5285.6}}&\cg{\fb{5514.5}}&\cg{5787.0}&\cg{5889.5}&\cg{6191.4}&\cg{5390.6}&3.5\\

		\hline
	\end{tabular}
  \begin{flushleft}
    {Note: The box indicates the favoured models, and the blue text indicates the second-ranked model in each column. Grey shades the models that are included in the 75\% MCS using the R method.}
\end{flushleft}
\end{table}

\setlength{\tabcolsep}{3pt}
\begin{table}[H]   
\centering
\footnotesize
\renewcommand{\arraystretch}{0.6}
	\caption{$2.5\%$ VaR \& ES joint loss function values across six indices.}
	\label{table: joint_25}
	\centering
	\begin{tabular}{p{1.8 in}cccccc|c}
		\hline
	\textbf{Model}  & \textbf{S\&P500} & \textbf{FTSE} & \textbf{NASDAQ}  & \textbf{  HSI  }  &  \textbf{ 
 DAX  } & \textbf{  SMI  } & \textbf{Avg Rank}   \\
		
        \rule{0pt}{15pt}\textbf{GARCH} &&&&&&& \\
        \hline 
		GARCH-t   &5274.1  & 5325.9 &  5810.0  &  5641.2 & 5871.3 & 5138.3  &  31.5\\
		
		EGARCH-t & 5187.0	& 5265.3 & 5810.0  &   5553.3 &  
5789.0 &  5070.5  &  26.0\\
		
		GJR-GARCH-t & 5207.1 & 5234.5 & 5762.1  &  5565.3 &  
5819.2 &5053.7  &   26.8 \\
		
		GARCH-QML-HS & 5373.5 & 5422.0 & 5900.4  & 5701.6 & 5962.6 &  5206.3 & 33.0  \\
		
		EGARCH-QML-HS & 5246.3 & 5314.7 & 5841.6  &   5590.7 &   
5824.0 & 5110.3  & 29.7  \\
		
		GJR-GARCH-QML-HS & 5273.4 & 5287.6 &  5811.0  &  5600.9 & 5867.4 & 5097.3  & 30.5 \\

     \rule{0pt}{15pt}\textbf{REGARCH-t} &&&&&&& \\
        \hline 
	  RV5  & 5183.7 & 5234.1 & 5523.0 & 5559.8 & 5866.3 &4994.9  & 25.3\\	
		RK & 4868.2 & 5220.0 & 5516.3 & \cellcolor[HTML]{C0C0C0}5513.0 & 5915.2 & 5031.9  & 26.2 \\
		
		BV & 5004.1 & 5219.9 & \cellcolor[HTML]{C0C0C0}5468.0 & 5566.4 & 5837.1 & 5027.7  & 23.2\\

		RV5-RK  & 4941.3 & 5192.0 &  \cellcolor[HTML]{C0C0C0}5443.2 & \cellcolor[HTML]{C0C0C0}5493.3 & 5849.0 & \cellcolor[HTML]{C0C0C0}4941.0  & 18.8 \\
				
		RV5-BV  & 4890.2 & 5212.4 &  \cellcolor[HTML]{C0C0C0}5453.4 & 5586.2 & 5837.4 & \cellcolor[HTML]{C0C0C0}4932.3  &20.5\\
				
		RK-BV  & 4870.3 & 5204.2 &  \cellcolor[HTML]{C0C0C0}5434.6 & \cellcolor[HTML]{C0C0C0}5490.1 &  5827.9 & \cellcolor[HTML]{C0C0C0}4948.9  & 16.7 \\
				
		RV5-RK-BV &4862.4 &  5180.0 &  \cellcolor[HTML]{C0C0C0}5435.3 & \cellcolor[HTML]{C0C0C0}5549.5 & 5819.4 & \cellcolor[HTML]{C0C0C0}4936.4  & 17.0\\

        \rule{0pt}{15pt}\textbf{ES-CAViaR} &&&&&&& \\
        \hline 
		ES-CAViaR-Add  &  5193.4 & 5198.2 & 5649.1 &  5558.2 &   5800.0 &  5094.8  &  25.2 \\

        \rule{0pt}{15pt}\textbf{ES-CAViaR-X}  &&&&&&& \\
        \hline 
        RV5  & 4995.6 & \cellcolor[HTML]{C0C0C0}5166.6 & 5455.1  & 5509.1 &  5762.9 &  4963.6  & 19.7 \\
		
		RK  & 5000.8 & \cellcolor[HTML]{C0C0C0}5169.6 &  5453.7  & \cellcolor[HTML]{C0C0C0}5509.2 &  5762.4 &  4990.6 &  20.2 \\
		
		BV  & 4947.5 & \cellcolor[HTML]{C0C0C0}5146.9&  5454.4 & \cellcolor[HTML]{C0C0C0}5507.8 & 5747.9 &  \cellcolor[HTML]{C0C0C0}4951.2 & 17.3 \\
		
        \rule{0pt}{15pt}\textbf{ES-X-CAViaR-X} &&&&&&& \\
        \hline 
		RV5  &  4905.8 &   \cellcolor[HTML]{C0C0C0}5146.3 &    \cellcolor[HTML]{C0C0C0}5382.0  &  \cellcolor[HTML]{C0C0C0}5494.9 &   5748.8 &  \cellcolor[HTML]{C0C0C0}4905.0  &  12.8 \\
		
		RK  &  4900.7 &  \cellcolor[HTML]{C0C0C0}5141.0 &  \cellcolor[HTML]{C0C0C0}5380.2  &  \cellcolor[HTML]{C0C0C0}5496.0 &   5742.2 &  4951.3  &  13.5\\
		
		BV  &  4900.5 &  \cellcolor[HTML]{C0C0C0}5117.4 & \cellcolor[HTML]{C0C0C0}5366.6  & \cellcolor[HTML]{C0C0C0}5491.2 &  5745.2 & \cellcolor[HTML]{C0C0C0}4923.0  & 10.7 \\
       
        \rule{0pt}{15pt}\textbf{Realized-ES-CAViaR} &&&&&&& \\
        \hline 
        RV5  &  4908.1 &  \cellcolor[HTML]{C0C0C0}5154.8 &   \cellcolor[HTML]{C0C0C0}5381.6 &  \cellcolor[HTML]{C0C0C0}5482.7 &   5731.0 &  \cellcolor[HTML]{C0C0C0}4945.1  & 12.5   \\
		
		RK  &  4949.7 &  \cellcolor[HTML]{C0C0C0}5159.4 &  \cellcolor[HTML]{C0C0C0}5389.6  &  \cellcolor[HTML]{C0C0C0}5488.7 & \cellcolor[HTML]{C0C0C0}5724.3 &  4981.7 &  14.8 \\
		
		BV  &  \cellcolor[HTML]{C0C0C0}4840.6 &  \cellcolor[HTML]{C0C0C0}5130.4 &  \cellcolor[HTML]{C0C0C0}5366.8 &  \cellcolor[HTML]{C0C0C0}5489.8 &  \cellcolor[HTML]{C0C0C0}5710.1  &   4959.1 & 9.0\\
               \rule{0pt}{15pt}\textbf{Log Realized-ES-CAViaR} &&&&&&& \\
        \hline 
        RV5  & 4898.7& \cellcolor[HTML]{C0C0C0}5151.0 &  \cellcolor[HTML]{C0C0C0}5376.4 &  \cellcolor[HTML]{C0C0C0}5487.3&   5740.4 &  \cellcolor[HTML]{C0C0C0}4910.1  & 10.5   \\
		RK  & 4904.6 & \cellcolor[HTML]{C0C0C0}5141.5 &  \cellcolor[HTML]{C0C0C0}5377.1 &  \cellcolor[HTML]{C0C0C0}5490.1 &  5739.0  &  4965.7   & 13.2   \\
		BV  & \cellcolor[HTML]{C0C0C0}4836.6&  \cellcolor[HTML]{C0C0C0}5113.6  &  \cellcolor[HTML]{C0C0C0}5364.1 &  \cellcolor[HTML]{C0C0C0}5500.7 &   5727.0 & \cellcolor[HTML]{C0C0C0}4924.4   & 8.3   \\
		\rule{0pt}{15pt}\textbf{Realized-ES-CAViaR-M} &&&&&&& \\
        \hline 
		RV5  &  4855.3 &\cellcolor[HTML]{C0C0C0}5102.8 & \cellcolor[HTML]{C0C0C0}\framebox[0.08\columnwidth]{5352.0}  &\cellcolor[HTML]{C0C0C0}5465.3 &  5720.4 &    \cellcolor[HTML]{C0C0C0}4895.0   & 4.5   \\
		RK  & 4873.3 & \cellcolor[HTML]{C0C0C0}5105.2 & \cellcolor[HTML]{C0C0C0}5360.0  & \cellcolor[HTML]{C0C0C0}\framebox[0.08\columnwidth]{5446.8} & \cellcolor[HTML]{C0C0C0}5709.3 &    \cellcolor[HTML]{C0C0C0}4908.9   & 5.5   \\
		BV  &  \cellcolor[HTML]{C0C0C0}\framebox[0.08\columnwidth]{4817.3} & \cellcolor[HTML]{C0C0C0}5091.5 & \cellcolor[HTML]{C0C0C0}5358.7  &\cellcolor[HTML]{C0C0C0}5462.9 & \cellcolor[HTML]{C0C0C0}5680.4 &   \cellcolor[HTML]{C0C0C0}4910.0   & {\color[HTML]{3166FF}3.5}  \\
		RV5-RK  & 4868.9 & \cellcolor[HTML]{C0C0C0}5110.0 &\cellcolor[HTML]{C0C0C0}5398.8  &\cellcolor[HTML]{C0C0C0}5459.4 &  5817.2 &   5002.9   & 13.2  \\
 RV5-BV  &  4853.6 &  \cellcolor[HTML]{C0C0C0}{\color[HTML]{3166FF}5077.2} & 5479.9  &   5543.9 & \cellcolor[HTML]{C0C0C0}5699.7  &  \cellcolor[HTML]{C0C0C0}4898.2 & 10.3 \\
		
 RK-BV  &   4837.9 &   5191.8 & \cellcolor[HTML]{C0C0C0}5382.2  & \ 5506.9&  \cellcolor[HTML]{C0C0C0}\framebox[0.08\columnwidth]{5608.0}  &   \cellcolor[HTML]{C0C0C0}{\color[HTML]{3166FF}4892.6} &  9.5  \\
		
 RV5-RK-BV  &  \cellcolor[HTML]{C0C0C0}{\color[HTML]{3166FF}4822.8} &   \cellcolor[HTML]{C0C0C0}\framebox[0.08\columnwidth]{5066.5} & \cellcolor[HTML]{C0C0C0}{\color[HTML]{3166FF} 5354.6}  &  \cellcolor[HTML]{C0C0C0}{\color[HTML]{3166FF}5455.7} & \cellcolor[HTML]{C0C0C0}{\color[HTML]{3166FF}5670.8}  &   \cellcolor[HTML]{C0C0C0}\framebox[0.08\columnwidth]{4877.7} &  \framebox[0.08\columnwidth]{1.7}  \\

		\hline
	\end{tabular}
  \begin{flushleft}
    {Note: The box indicates the favoured models, and the blue text indicates the second-ranked model in each column. Grey shades the models that are included in the 75\% MCS using the R method.}
\end{flushleft}
\end{table}

\setlength{\tabcolsep}{3pt}
\mbox{}\vspace*{0.3cm} 
\begin{table}[H]    
\centering
\footnotesize
\renewcommand{\arraystretch}{0.6}
	\caption{75\% model confidence set with R and SQ methods on 1\% and 2.5
 \% probability levels. } 
	\label{table: MCS}
	\centering
	\begin{tabular}{p{1.8 in}cccc|cccc|c}
 		\hline
		  & \multicolumn{4}{c}{1\%}  & \multicolumn{4}{c}{2.5\%} \\
		\textbf{Model}  & \multicolumn{2}{c}{Quantile} & \multicolumn{2}{c}{Joint} & \multicolumn{2}{c}{Quantile} & \multicolumn{2}{c}{Joint} &  \\
		\hline
		& \textbf{R } & \textbf{SQ } & \textbf{R } & \textbf{SQ }& \textbf{R } & \textbf{SQ } & \textbf{R } & \textbf{SQ } & \textbf{Total}\\

        \rule{0pt}{15pt}\textbf{GARCH} &&&&&&&& \\
        \hline 
GARCH-t   &2&2&2&2&0&0&0&0&8\\
EGARCH-t &2&3&2&2&3&3&0&1&16\\
GJR-GARCH-t &4&4&3&1&3&4&0&1&20\\
GARCH-QML-HS&2&2&3&2&0&0&0&0&9\\
EGARCH-QML-HS &5&5&3&3&3&1&0&1&21\\ 
GJR-GARCH-QML-HS&4&4&3&3&3&1&0&0&18\\
     \rule{0pt}{15pt}\textbf{REGARCH-t} &&&&&&&& \\
        \hline 
RV5&2&3&0&1&1&1&0&1&9\\
RK &3&4&3&2&3&3&1&2&21\\
BV &2&3&4&2&3&1&1&2&18\\
RV5-RK  &4&4&3&3&5&5&3&4&31\\
RV5-BV  &3&5&2&3&3&3&2&4&25\\
RK-BV  &4&5&3&4&5&5&3&4&33\\
RV5-RK-BV &4&6&4&4&5&5&3&5&36\\
        \rule{0pt}{15pt}\textbf{ES-CAViaR} &&&&&&&& \\
        \hline 
ES-CAViaR-Add&2&2&2&2&0&0&0&1&9\\
        \rule{0pt}{15pt}\textbf{ES-CAViaR-X}  &&&&&&&& \\
        \hline 
RV5&5&5&4&5&1&1&1&2&24\\
RK &4&5&4&4&2&0&2&2&23\\
BV &5&5&4&5&2&2&3&4&30\\
        \rule{0pt}{15pt}\textbf{ES-X-CAViaR-X}  &&&&&&&& \\
        \hline 
RV5&5&5&4&5&6&4&4&5&38\\
RK &4&4&5&4&5&5&3&5&35\\
BV &5&6&5&6&5&5&4&5&41\\
        \rule{0pt}{15pt}\textbf{Realized-ES-CAViaR} &&&&&&&& \\
        \hline 
RV5&5&5&4&5&5&3&4&5&36\\
RK &5&5&4&4&5&4&4&3&34\\
BV &6&6&5&6&5&4&5&5&42\\
        \rule{0pt}{15pt}\textbf{Log-Realized-ES-CAViaR} &&&&&&&& \\
        \hline 
RV5&5&6&4&6&6&5&4&5&41\\
RK &5&5&5&5&5&4&3&4&36\\
BV &6&6&5&6&6&5&5&5&44\\
        \rule{0pt}{15pt}\textbf{Realized-ES-CAViaR-M} &&&&&&&& \\
        \hline 
RV5&5&5&4&5&6&5&4&5&39\\
RK &5&5&5&6&6&6&5&5&43\\
BV &6&6&5&6&6&6&6&5&\cb{46}\\
RV5-RK  &5&6&4&5&4&4&3&4&35\\
RV5-BV  &6&6&6&6&6&5&3&4&42\\
RK-BV  &6&6&6&6&5&5&3&6&43\\
RV5-RK-BV &6&6&6&6&6&6&6&6&\fb{48}\\

		\hline
	\end{tabular}
 \begin{flushleft}
    {Note: The box indicates the favoured models, and the blue text indicates the second-ranked model based on the Total column.}
\end{flushleft}
 \end{table}

\section{Conclusion}
This paper proposes a new semi-parametric joint VaR and ES forecasting framework incorporating multiple realized measures. The proposed Realized-ES-CAViaR-M models generate highly competitive risk forecasting results regarding quantile loss, VaR and ES joint loss, and MCS backtest. In particular, the proposed Realized-ES-CAViaR-M models produce favourable results compared to their parametric counterpart, e.g., the REGARCH model, and the semi-parametric counterparts, e.g., ES-X-CAViaR-X and Realized-ES-CAViaR. 

This work can be improved by considering more realized measures, including their sub-sampled versions and different frequencies. Moreover, the proposed framework includes single lags only, which can be extended to multiple lags. Finally, alternative versions of the model, such as a multiplicative time-varying relationship between VaR and ES via incorporating the information from multiple realized measures, could be considered in future work. 

\section*{Disclosure of Interest}
No conflict of interests to be declared.

\section*{Disclosure of Funding}
No funding was received.

\section*{Data Availability Statement}
Data were downloaded from Oxford-man Institute’s realized library. The authors confirm that the data supporting the findings of this study are available within the supplementary materials of the paper.

\clearpage 

\bibliographystyle{apalike}
 \bibliography{referencelist}
\end{document}